\documentclass{aa}  
\usepackage{graphicx}
\usepackage{txfonts}
\usepackage{lipsum}
\usepackage{float}
\usepackage[dvipsnames]{xcolor}
\usepackage{cancel}
\usepackage[normalem]{ulem}
\usepackage{soul}
\usepackage{subcaption}         
                                
\usepackage{lscape}            
                               
\usepackage{placeins}           
                                
\usepackage{natbib}
\bibpunct{(}{)}{;}{a}{}{,}

\usepackage[normalem]{ulem} 
\usepackage{hyperref}
\hypersetup{
    colorlinks,
    linkcolor={blue},
    citecolor={blue},
    urlcolor={blue}
}

\begin{document}

\title{Unveiling the nature of barium stars} 

   \subtitle{I. Asteroseismic masses and the evolutionary link between Ba dwarfs and giants}

\author{Lupamudra Sarmah\inst{1,2}\fnmsep\thanks{Corresponding author: lupamudra.sarmah@iiap.res.in}
        \and Yerra Bharat Kumar \inst{1}
        \and Simon W. Campbell\inst{3}
         \and Sunayana Maben\inst{4,1}
        \and Bacham E. Reddy\inst{5,1}
        }

   \institute{Indian Institute of Astrophysics, 100ft road, Koramangala, Bangalore, 560034, India
   \and Pondicherry University, R.V. Nagar, Kalapet 605014, Puducherry, India
   \and School of Physics and Astronomy, Monash University, Clayton, Victoria, Australia
   \and CAS Key Laboratory of Optical Astronomy, National Astronomical Observatories, Chinese Academy of Sciences, Beijing 100101, People's Republic of China
   \and Department of Physics, Indian Institute of Technology Jammu, Jammu 181221, India}

   \date{}

  \abstract 
   {Barium star systems are excellent sites for studying asymptotic giant branch (AGB) nucleosynthesis, binary evolution, and
mass transfer processes. However,  an accurate estimation of their fundamental stellar parameters is still
lacking.}
{We measure accurate and precise masses of Ba stars using asteroseismology. This enables us to constrain the nature, origin, and evolution of these binary systems.}
   {Using data from the Transiting Exoplanet Survey Satellite, we made the first extensive asteroseismic mass measurements of Ba stars. Our sample comprises 31 Ba giants and 13 Ba dwarfs.
   For some, we were able to measure $\Delta P$, thereby ascertaining their evolutionary phase. 
   With reliable asteroseismic masses, we then constructed a grid of stellar models across the relevant mass range, where we accreted AGB material using composition from existing yields.}
   {We found that the average masses of the Ba dwarfs and Ba giants are significantly different ($1.29\pm0.09~\rm{M}_\odot$ versus $1.96\pm0.16~\rm{M}_\odot$, respectively; with typical individual mass uncertainties of $\sim10\%$). However, their mass distributions peak at about the same mass ($\sim1.3~\rm{M}_\odot$). While our sample of Ba giants spans the low- and intermediate-mass regime, we found no intermediate-mass Ba dwarfs. The abundance trends of $s$-process elements ([s/Fe], [hs/Fe], and [ls/Fe]) show an overall anti-correlation with stellar mass, particularly in the low-mass regime ( $<2~\rm{M}_\odot$ ), for giants and dwarfs. The stellar models adopting Monash AGB yields can satisfactorily reproduce the observed light elements, $s$, and heavy-$s$ abundance trends simultaneously, with an accreted mass of $0.1-0.5~\rm{M}_\odot$ for the majority of the Ba stars. 
   However, the models fail to explain the light-$s$ abundances and, consequently, the [hs/ls] ratio. We found that most Ba stars had AGB companions in the mass range $1-4$~M$_\odot$.}
   {
   Our results support an evolutionary scenario in which Ba giants evolve from Ba dwarfs, with mass accretion occurring while the progenitor Ba star is still on the main sequence. In this scenario, a substantial number of intermediate-mass Ba dwarfs are expected. We argue that they remain undetected due to observational bias. We found that post-accretion additional mixing in our models is critical to explain the observed $s$-process abundances in Ba dwarfs and the low C isotopic ratio ($<30$) in Ba giants. The mismatch between the model and the observed [hs/ls] ratio suggests that the chemical enrichment of Ba stars cannot be explained by standard single-star AGB yields alone. This may be due to (i) their binary nature altering AGB evolution, (ii) missing or modified nucleosynthesis processes in AGB models, or (iii) additional sources of pollution. }


   \maketitle
\nolinenumbers

\section{Introduction}
Barium (Ba) stars are a subclass of slow neutron-capture process ($s$-process) enhanced stars in the near-solar metallicity regime. First identified by~\cite{1951ApJ...114..473B}, these stars exhibit unusually strong spectral lines of $s$-process elements, particularly Ba II ($4554$ \r{A}) and Sr II ($4077$ \r{A}), along with molecular bands of CH, CN, and C$_2$. One of the major sites for the $s$-process nucleosynthesis is thought to be the interiors of asymptotic giant branch (AGB) stars, where these elements are synthesised as a result of neutron-capture processes at low neutron densities with $N_n\sim10^8$~cm$^{-3}$~\citep{2011RvMP...83..157K,2014PASA...31...30K,2023ARNPS..73..315L}. These elements are then brought to the stellar surface by third dredge-up (TDU) episodes during the thermally pulsing AGB (TP-AGB) phase, making AGB stars intrinsically enriched in $s$-process elements. The observation of $s$-process enhancements in Ba stars posed a fundamental challenge to single-star evolution and nucleosynthesis theory, as these stars are not evolved enough to have undergone the TDU. Classically, Ba stars were thought to occupy the red giant branch (RGB) and pre-AGB phase of the Hertzsprung-Russell (HR) diagram, called Ba giants. The mystery of their origin was finally resolved with the evidence of binarity: detections of UV excess and radial velocity monitoring revealed that most Ba stars are in binary systems with a white dwarf (WD) companion, and this detection also established the mass transfer hypothesis as the most probable formation channel~\citep{1980ApJ...238L..35M,1984ApJ...278..726B,1990ApJ...352..709M,2019A&A...626A.127J, 2019A&A...626A.128E, 2023A&A...671A..97E}. According to this hypothesis, the $s$-process elements were transferred to the Ba star progenitor from a more massive companion during its AGB phase (now a WD). Since they retain the chemical signatures of their companions, Ba star systems are excellent sites for probing AGB nucleosynthesis and binary interactions.

Over the past few decades, detailed high‑resolution spectroscopic studies have refined our knowledge of these stars, particularly Ba giants. The criteria for classifying these objects are not universally defined, however. For example, ~\cite{2016MNRAS.459.4299D} adopted a criterion of [s/Fe]\footnote{where [s/Fe] is the mean abundance of the $s$-process elements} $\geq0.25$~dex for a star to be considered as a Ba star and provided detailed abundances based on high-resolution spectroscopy for an extensive sample of 182 Ba giants and candidates. Their analysis revealed that $\alpha$-elements and iron-peak elements in Ba giants show similar abundances to those of field stars. Furthermore, the authors noted an anti-correlation between [s/Fe] and [hs/ls]\footnote{[hs/ls]=[hs/Fe]-[ls/Fe], where [hs/Fe], and [ls/Fe] are the average abundances of heavy-$s$ ($hs$), and light-$s$ ($ls$) process elements respectively.} with metallicity. This was attributed to greater neutron exposure per seed nuclei in low-metallicity AGB stars, leading to an increased production of $hs$-elements compared to $ls$-elements. The extensive chemical abundance analysis of light and heavy elements in Ba giants by~\cite{2018A&A...618A..32K} and \cite{2020MNRAS.492.3708S} showed that the chemical signatures are consistent with the operation of a $^{13}$C$(\alpha,n)^{16}$O neutron source in $2-4~\rm{M}_\odot$ AGB companions.

Recently, Ba stars have been detected in more unevolved phases of evolution, such as the main sequence (MS) and subgiant (SG) branch, called Ba dwarfs~\citep{1994A&A...281..775N,2006A&A...454..895A,2018MNRAS.474.2129K, 2019A&A...626A.128E}. These are relatively rare and therefore less well studied. Ba dwarfs are particularly useful as they retain the chemical signature of their AGB companions and are unaffected by the subsequent convective mixing during the first dredge-up (FDU) in giants. Moreover, their abundance patterns can provide clues on additional stellar mixing in the MS and SG phase~\citep{2009PASA...26..176H}. Using high-resolution spectroscopy,~\cite{2024AJ....167..184R} recently discovered three new Ba dwarfs. They also found that low-mass AGB models ($\leq3~\rm{M}_\odot$) with a dilution factor of $\leq0.45$ effectively reproduced the observed abundances of their Ba dwarfs. More recently, machine-learning approaches have been used to identify the mass and metallicity of AGB companions that polluted the now Ba star based on their observed $s$-process element abundances.~\cite{2023A&A...672A.143D} used neural networks and a nearest-neighbour algorithm to recognise abundance patterns in Ba stars and then determined the mass and metallicity of the AGB model that predicted the surface abundance closest to that of the observed Ba star values. Their result yielded an average AGB mass of $2.23-2.34~\rm{M}_\odot$ as the probable polluter of their Ba star sample. Additionally,~\cite{2024A&A...688A.164V} used a random forest method to compare the observed Ba star abundances with those of AGB models and investigated whether the AGB models systematically under- or overpredict the abundances.
 
While much work has been done to characterise the chemical peculiarities of Ba stars, an accurate estimation of their fundamental stellar parameters is still lacking. Traditionally, previous studies have estimated Ba star masses from their positions on the HR/Kiel diagram, which is subject to large uncertainties. According to these estimates, Ba giants have a mass range of $1-6~\rm{M}_\odot$ with typical uncertainties of $\sim 30-60 \%$, or even larger~\citep{2016MNRAS.459.4299D}. Using a similar technique,~\cite{2019A&A...626A.128E} found that the mass distribution of Ba dwarfs peaks at a much lower mass than giants, suggesting that Ba dwarfs are unlikely to represent the direct progenitors of Ba giants. Significant uncertainty still surrounds the stellar parameters determined for these stars. For instance, these mass estimates are very sensitive to the adopted metallicities and model assumptions of the stellar tracks, such as the mixing-length parameter and overshoot. Moreover, evolutionary tracks of different masses overlap in the RGB phase and can also overlap with the red clump (core helium-burning, CHeB) and early-AGB phases. Thus, there are multiple sources of degeneracies that make mass determinations difficult. Overcoming these uncertainties and obtaining reliable stellar parameters is crucial because it might provide a coherent picture of the formation history and chemical trends of Ba stars. Precise masses, radii, and surface gravity can place robust constraints on the accretion history and subsequent mixing of the accreted AGB material in the envelope of the Ba star. This in turn would also help us to link the observed chemical patterns to their stellar parameters and to nucleosynthesis signatures in their corresponding AGB companions. 

Asteroseismology has revolutionised our ability to probe the interior of stars by studying their intrinsic oscillations~\citep{2013ARA&A..51..353C,2021RvMP...93a5001A}. It allows us to determine accurate and precise stellar parameters, including mass and radius, together with their evolutionary phase, thereby removing the long-standing degeneracies and uncertainties surrounding these parameters. 
The combination of asteroseismic masses with spectroscopic abundances offers a unique opportunity to investigate how stellar structure and evolution affect the chemical anomalies in chemically peculiar stars. 

We advance previous efforts of mass determinations by deriving asteroseismic masses and evolutionary phases for a sample of 31 Ba giants and 13 Ba dwarfs using data from the Transiting Exoplanet Survey Satellite (TESS;~\citealt{2015JATIS...1a4003R}). This enabled us to investigate the mass distributions and behaviour of the $s$-process and light elements with the stellar mass of Ba stars. Further, by combining these asteroseismic estimates with stellar models, we explore the role of internal mixing and other factors that shape their light- and heavy-element abundance patterns.

\section{Sample selection and methods}
\label{Methods and Results}
\subsection{Sample selection}
We created a comprehensive catalogue of all known Ba giants and dwarfs reported in the literature as of March 2024, comprising 436 Ba giants and 80 Ba dwarfs. This includes systems with and without detailed spectroscopic abundance analysis. We then cross-matched this catalogue with the targets observed by the TESS and Kepler~\citep{2010Sci...327..977B} missions in order to identify those suitable for asteroseismic analysis. This step yielded 410 Ba giants and 75 Ba dwarfs that have been observed at least once by TESS, while 8 Ba giants and 1 Ba dwarf were identified in the K2 field. 

For uniformity and data quality, we focused exclusively on TESS data products processed by the Science Processing Operations Centre (SPOC; \citealt{2016SPIE.9913E..3EJ}), which provide high-quality light curves (LCs) while correcting for systematics. This criterion resulted in 266 Ba giants and 73 Ba dwarfs. 
We downloaded all available SPOC LCs and examined each power spectrum visually for the presence of a clear power excess (PE). Stars that lacked detailed, high-resolution spectroscopic abundances, did not show any PE, or had strong contamination or instrument artefacts were discarded. This process resulted in a final working sample of 31 Ba giants ($7\%)$ and 13 Ba dwarfs ($16\%$) suitable for asteroseismic analysis.

\subsection{Asteroseismic data}
We downloaded the SPOC LCs from the Mikulski Archive for Space Telescopes (MAST\footnote{\url{https://mast.stsci.edu/portal/Mashup/Clients/Mast/Portal.html}}) website using the Python package Lightkurve~\citep{2018ascl.soft12013L}. In addition to the standard SPOC products derived from two-minute cadence target pixel files, we also utilised TESS-SPOC data~\citep{Caldwell2020}, which extends the SPOC pipeline processing to generate LCs from full-frame images (FFIs). For our analysis, we adopted the Pre-search Data Conditioning Simple Aperture Photometry (PDCSAP) flux, which is derived from the original Simple Aperture Photometry (SAP) flux by removing long-term instrumental and systematic trends through the application of co-trending basis vectors.
The resulting PDCSAP flux is cleaner, with a better signal-to-noise ratio (S/N) than the original SAP flux, and is thus suitable for asteroseismology. We restrict our analysis to LC measurements with a quality flag of zero~\citep{Vanderspek2018} to ensure the best quality data points with no known instrumental or processing issues. This criterion removed only a negligible fraction (an average of $0.6\%$) of data points. 
Subsequently, we remove outliers by applying a five sigma-clipping using the remove$\_$outliers function in Lightkurve~\citep{2024ApJS..271...17Z}. Finally, all corrected LCs from different sectors were normalised to place them on a consistent scale and remove systematics. We then stitched them together, producing a continuous time series for asteroseismic analysis (e.g.~\citealt{2011MNRAS.414L...6G}).

\subsection{Measuring global asteroseismic parameters}
The flux variation recorded in the LCs with time is a superposition of intrinsic stochastic oscillations, convective background and instrumental noise~\citep{2011ApJ...741..119M,2014A&A...570A..41K, 2017A&A...605A...3C}.
We convert the LCs to power spectral density (PSD) using the Lomb-Scargle method~\citep{1976Ap&SS..39..447L,1982ApJ...263..835S,2018ApJS..236...16V}, owing to the unevenly spaced observations by TESS. In the PSD, the stochastic oscillations manifest themselves as a Gaussian PE above the background, as shown in Figure~\ref{psd}.

\subsubsection{Estimating \texorpdfstring{$\nu_{max}$}{nu_max} and its uncertainties}
\label{estimating numax}
The frequency of maximum oscillation, $\nu_{max}$, scales with the acoustic cutoff frequency of the star and hence provides important constraints on the mass and gravity of the star~\citep{1991ApJ...368..599B}. To estimate $\nu_{max}$, we first model the PSD by employing a function, $P(\nu)$, consisting of a Gaussian envelope for the PE, super-Lorentzian components for granulation background and a flat line for the constant white noise \citep{2014A&A...570A..41K, 2017A&A...605A...3C},
\begin{equation}
    P(\nu)=W + R(\nu) \left[ G(\nu) + B(\nu)  \right],
\label{eq1}
\end{equation}
where $W$ is the white noise, dominant at high frequencies, $G(\nu)$ is the Gaussian envelope function that is used to model the PE,
\begin{equation}
    G(\nu)= H_g \exp \left[ -\frac{(\nu - \nu_{max})^2}{2 \sigma^2}\right],
\end{equation}
with $H_g$, $\nu_{max}$, and $\sigma$ being the height, central frequency, and width of the Gaussian envelope. The granulation background is modelled by two super-Lorentzian components,
\begin{equation}
    B(\nu)= \sum_{i=1}^2 \frac{\zeta a_i^2/b_i}{1+(\nu/b_i)^4},
\label{eq2}
\end{equation}
where $a_i$ is the rms amplitude, $b_i$ is the characteristic frequency of the $i^{th}$ background component, and $\zeta=2\sqrt{2}/\pi$ is the normalisation factor~\citep{2013ApJ...767...34K,2014A&A...570A..41K}. The two super-Lorentzian components in this work characterize meso-granulation ($a_1,b_1$) and granulation ($a_2,b_2$) with $a_1 > a_2$, and $b_1 < b_2$. It has been well established that these amplitudes and timescales are strongly correlated  with $\nu_{max}$~\citep{2017A&A...605A...3C},
\begin{equation}
    a \propto \nu_{max}^{-1/2}~,~b\propto \nu_{max}.
\label{eq3}
\end{equation}
To account for the attenuation caused by the finite sampling integration time of the telescope, the components of granulation and the Gaussian envelope are  modulated by a response function,
\begin{equation}
    R(\nu)=\mathrm{sinc}^2\left(\frac{\pi \nu}{2 \nu_{Nyq}}\right),
\label{eq3a}
\end{equation}
where $\nu_{Nyq}=(2\Delta t)^{-1}$ is the Nyquist frequency, with $\Delta t$ being the cadence of observation. It should be noted that since white noise is frequency independent, it is unaffected by sampling effects and is therefore not modulated by $R(\nu)$.

We employ the Bayesian technique using Markov chain Monte Carlo (MCMC;~\citealt{2013PASP..125..306F}) to estimate $\nu_{max}$ and its error by fitting Eq.~\ref{eq1} to the observed PSD. For PSDs with more than 25000 data points, we binned the periodogram to reduce the number of data points for faster computation while preserving the overall power distribution.
For example, we considered a bin size of $0.1~\mu$Hz for the star HD32712, which has nine sectors of data available.

\begin{figure}
    \centering
    \includegraphics[width=\hsize]{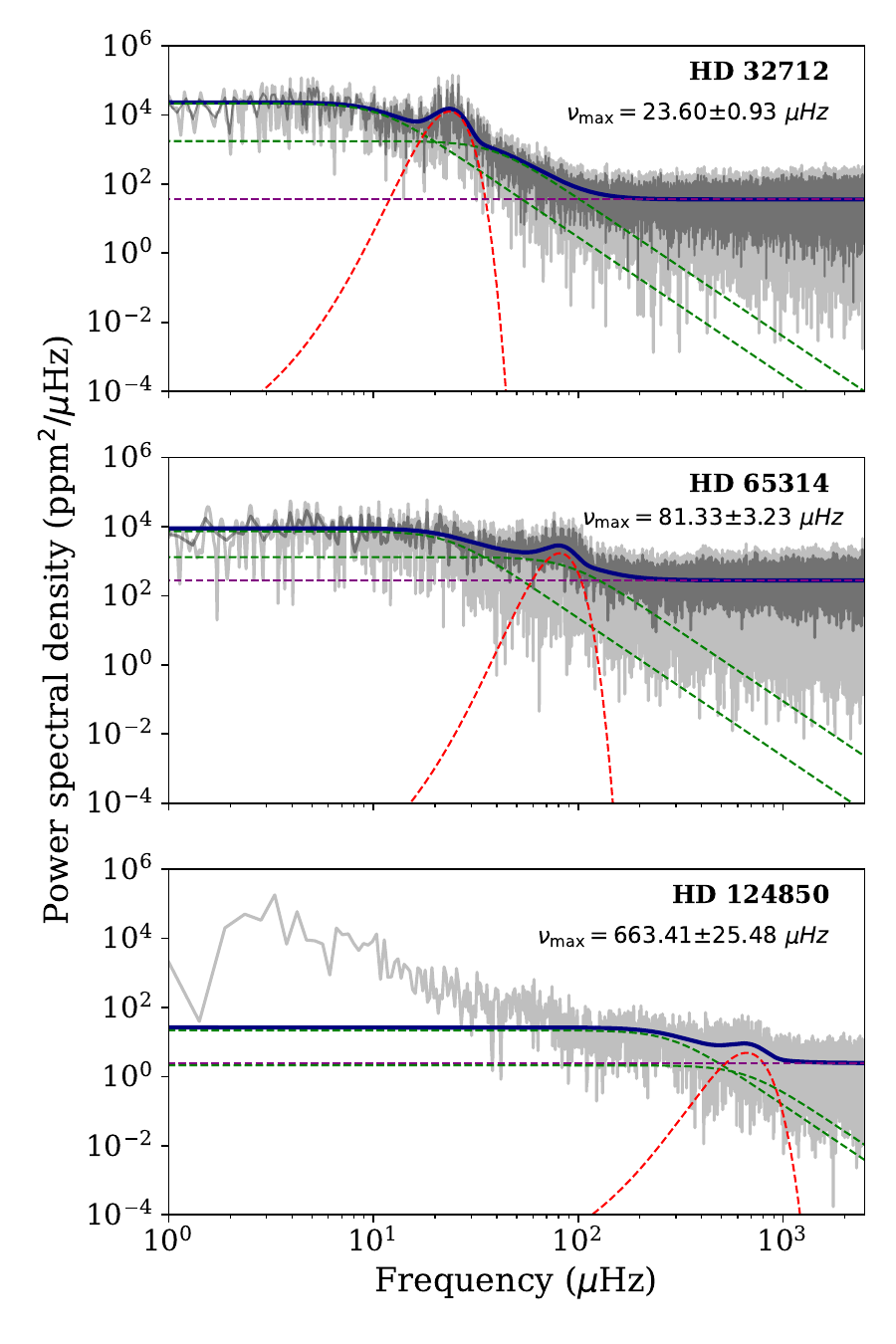}
    \caption{Representative PSDs of sample stars spanning a range of $\nu_{max}$ and the corresponding background fitting (Eq.~\ref{eq1}). The grey data points show the observed PSD, and the dark grey data points show the binned data. The solid navy line shows the best fit. The dashed red line represents the Gaussian component, the dashed green lines denote the super-Lorentzian components, and the dashed purple lines show the white noise.}
    \label{psd}
\end{figure}

The initial guess for $a_i$ and $b_i$ is estimated using Eq.~\ref{eq3}, while preliminary estimates for $\nu_{max},~H_g,~\sigma$ are obtained by visually inspecting the PE. The MCMC sampling is performed for 500 burn-ins, 200 walkers, and at least 2000 steps, and finally, $\nu_{max}$ is estimated as the median of the posterior probability distribution. We show a few representative examples of the results of the fitting procedure in Figure~\ref{psd}.

To estimate uncertainties in $\nu_{max}$, we adopt a procedure similar to that of pySYD\footnote{\url{https://github.com/ashleychontos/pySYD}}~\citep{2022JOSS....7.3331C}, an open source Python version of the
widely tested IDL-based SYD pipeline~\citep{2009CoAst.160...74H} for estimating global asteroseismic parameters. We perturb the observed PSD 100 times using a $\chi^2$ distribution having two degrees of freedom, and for each perturbed PSD, we repeat the background fitting and $\nu_{max}$ determination independently. Finally, the standard deviation of the resulting converged $\nu_{max}$ values is taken as the uncertainty. The estimated $\nu_{max}$ values and their uncertainties for Ba giants and dwarfs are given in Table~\ref{gba dba param}. To cross-check our $\nu_{max}$ estimations, we independently estimated $\nu_{max}$ values of our sample stars using pySYD and found strong agreement between the values derived from the two methods (see Appendix~\ref{validation} and Figure~\ref{numax compare}). 

We found that all of our  Ba giants showed clear detection of solar-like oscillations. Among the dwarfs, we found eight of them with clear detection of PE ($62 \%$ of the Ba dwarf sample), whereas the remaining five show marginal detection of PE and hence uncertain $\nu_{max}$.
Further observations in additional sectors would improve the S/N for reliable measurement of their global asteroseismic properties.
The $\nu_{max}$ detection status of Ba dwarfs is summarised in Table~\ref{gba dba param}.

\subsubsection{\texorpdfstring{Estimating $\Delta\nu$}{Estimating Delta nu}}
\label{estimating delnu}
According to the asymptotic theory~\citep{1980ApJS...43..469T}, the $p$-modes of a given degree are regularly spaced in frequency, characterised by the large frequency separation ($\Delta\nu$). When combined with $\nu_{max}$, it is widely used to estimate stellar mass and radius. We measure $\Delta\nu$ using pySYD, which uses an  autocorrelation function (ACF) method applied to the background‑corrected power spectrum~\citep{2009CoAst.160...74H,2022JOSS....7.3331C}. In short, pySYD uses a background function similar to Eq.~\ref{eq1} to model the PSD and estimates $\nu_{max}$ from the peak of the smoothed PE. After subtracting the background model from the PSD, the background-corrected PSD is used to calculate the ACF, collapsed over all frequency spacing. It identifies the peak in this ACF closest to the expected $\Delta\nu$ and fits a Gaussian to it, with the Gaussian peak providing the $\Delta\nu$. Figure~\ref{pysyd} shows the pySYD analysis for the Ba giant HD30554, illustrating the determination of $\nu_{max}$ using the smoothed PSD and $\Delta\nu$ using the ACF method.

We could only estimate $\Delta\nu$ confidently for 14 Ba giants and 5 Ba dwarfs, as given in Table~\ref{delnu mass}. This limitation is due to the relatively short observational duration and the consequently low frequency resolution of the data. 
 Additionally, elevated noise levels in the PSD can compete and obscure the oscillation signal, preventing the ACF method from identifying a clear $\Delta\nu$ peak.

\subsubsection{\texorpdfstring{Estimating $\Delta P$ for evolutionary phase}{Estimating ΔP for evolutionary phase}}
Dipole mixed modes ($l=1$) observed in the PSD of red giant (RG) stars provide a powerful tool for probing the internal structure of the star and have been used to distinguish hydrogen shell burning giants from CHeB stars~\citep{2011Natur.471..608B,2016A&A...588A..87V}. This distinction can be made by measuring the period separation between consecutive dipole mixed modes, known as average period spacing ($\Delta P$). To estimate $\Delta P$, we first divide the observed PSD by the background function, while considering only the white noise and granulation background terms in Eq.~\ref{eq1}. This gives us an idea of the S/N of the modes. Because TESS data has low S/N due to its short observational time, we have only selected those frequencies with S/N$~>3$ for the detection of significant $l=1$ modes~\citep{2019MNRAS.487..782L,2024ApJS..271...17Z}. Next, we identify at least four regions in the PSD containing $l=1$ modes, which are typically located between $l=0$ and $l=2$ modes. Finally, we estimate the period between consecutive $l=1$ modes and the average value is adopted as $\Delta P$, as shown in Figure~\ref{delp}. The standard error of the mean is taken as the uncertainty on $\Delta P$. Since we measure the average $\Delta P$ instead of asymptotic period spacing, we consider a threshold of $\Delta P \geq 150~$s for CHeB stars~\citep{2013ApJ...765L..41S}. The estimated $\Delta P$ and their associated uncertainties for the seven Ba giants of our sample are given in Table~\ref{delnu mass}.
\label{evs}

\subsection{Luminosities}
Combining stellar luminosities $(L)$ with global asteroseismic properties can help in estimating stellar masses using the scaling relations as explained in Sect.~\ref{scaling rel}. We estimated $L$ of Ba giants and dwarfs by combining astrometric, photometric, and spectroscopic data, using the standard formula
\begin{equation}
    \log{(L/L_\odot)}=\left(M_{bol,\odot}-M_V-BC+A_V \right)/2.5,
    \label{eq6}
\end{equation}
where, $M_{bol,\odot}=4.74$~mag is the absolute bolometric magnitude of the Sun,  $M_V$ is the absolute $V$-band magnitude of the star, $BC$ is the bolometric correction, and $A_V$ is the interstellar extinction in the $V$-band. The absolute magnitude is given by
\begin{equation}
   M_V= V - 5\log{d} + 5,
\end{equation}
 where $V$ is the apparent visual magnitude and $d$ is the distance in parsecs. Distances and their uncertainties were taken from the catalogue of~\cite{2021AJ....161..147B}, where we adopted the photogeometric distances. 
  We estimated the  $V$ magnitude and its uncertainty using the colour–colour transformations provided by~\cite{2021A&A...649A...3R}, which relate Gaia $G$ magnitudes to the Johnson-Cousins photometric system.
 
To correct for interstellar extinction, $A_V$, we used the three-dimensional dust map of reddening by~\cite{2019ApJ...887...93G} which is based on Gaia parallaxes and photometry from  Pan-STARRS 1~\citep{2016arXiv161205560C} and 2MASS~\citep{2006AJ....131.1163S}. However, this map only covers the sky north of a declination ($\delta$) of $-30^\circ$. Thus, for the stars with $\delta < -30^\circ$ we used the extinction law given by~\cite{1998A&A...336..137C} to estimate the extinction and its errors. For giants, bolometric corrections and their errors are determined from the empirical relations given in Eqs. 17 and 18 of ~\cite{1999A&AS..140..261A}, which take into account effective temperature ($T_{\rm eff}$) and metallicity ([Fe/H]). For dwarfs, we use the empirical bolometric correction law based on $T_{\rm eff}$ of the star provided in Table 1 of~\cite{2010AJ....140.1158T}. We have collected $T_{\rm eff},$~[Fe/H], and their corresponding errors for the sample stars from spectroscopic studies available in the literature, and their adopted values are given in Table~\ref{gba dba param}, along with their references.

\begin{figure*}
  \includegraphics[width=0.9\hsize]{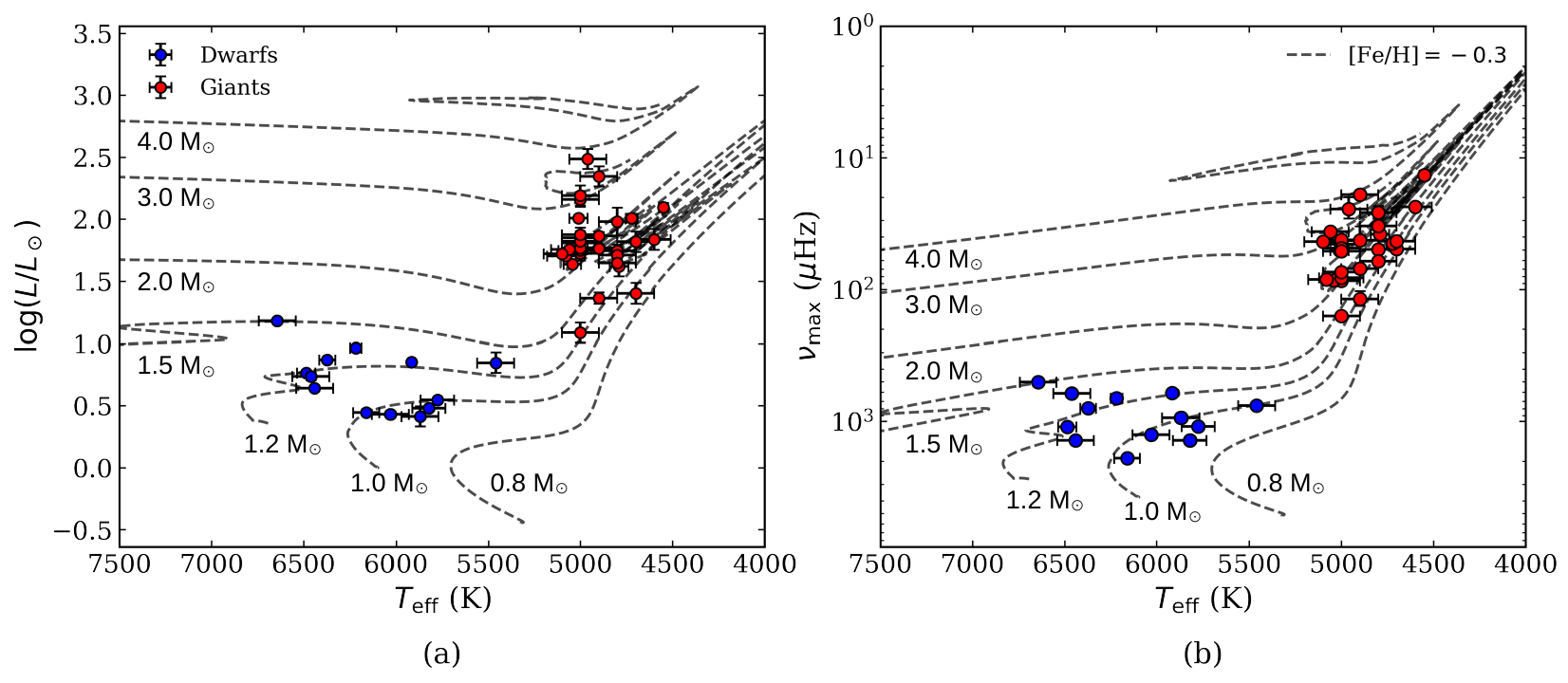}
     \caption{\textit{Panel (a)}: HR diagram of the sample Ba giants (red) and Ba dwarfs (blue). \textit{Panel (b)}: Same but on the $\nu_{max}$--$T_{\rm eff}$ plane. Evolutionary tracks of $0.8-4~\rm{M}_\odot$ and [Fe/H]$~=-0.3$ taken from the  MIST database~\citep{2016ApJS..222....8D,2016ApJ...823..102C} are shown as dashed grey lines.}
     \label{hrd_numax_tracks}
\end{figure*}

Finally, to estimate the errors in $\log{(L/L_\odot)}$, we propagated the errors in all the contributing quantities in Eq.~\ref{eq6}. The estimated $L$ are given in Table~\ref{gba dba param}. Using the estimated $L$, we show the location of our sample Ba giants and dwarfs on the HR diagram in Figure~\ref{hrd_numax_tracks}a. In Figure \ref{hrd_numax_tracks}b, these stars are overplotted on the $\nu_{max}$--$T_{\rm eff}$ plane using our measured $\nu_{max}$ values. We have used evolutionary tracks of [Fe/H]~$=-0.3$, which is typical for Ba stars (see Table~\ref{gba dba param}). The Ba dwarfs are clearly in the MS and SG phases, whereas the Ba giants span across the RGB and CHeB phases. However, many of the Ba giants occupy a similar range in $L$, and $\nu_{max}$ where the RGB and CHeB phase of different evolutionary tracks overlap, making it difficult to distinguish between Ba giants in these two evolutionary phases, or estimate accurate masses from these diagrams.

\subsection{Scaling relations and estimating masses}
\label{scaling rel}
The global asteroseismic parameters provide strong diagnostics to estimate stellar parameters via scaling relations. It has been shown by~\cite{1991ApJ...368..599B} that $\nu_{max}$ scales with the acoustic cutoff frequency and is related to stellar parameters by $\nu_{max} \propto g T_{\rm eff}^{-1/2}$, where $g$ is the surface gravity of the star. Meanwhile, $\Delta\nu$ is related to the acoustic diameter and thus scales with the mean density ($\rho$) of the star, $\Delta\nu \propto \rho^{1/2} $~\citep{1986ApJ...306L..37U}. Combining these relations with the Stefan-Boltzmann law, $L\propto R^2 T_{\rm eff}^4$, where $R$ is the radius of the star, we can derive the following scaling relations for the asteroseismic mass~\citep{2009ApJ...700.1589S,2010A&A...522A...1K,2016MNRAS.461..760M,2022MNRAS.515.3184H}:
\begin{equation}
    \left(\frac{M}{M_\odot}\right) \approx 
    \left(\frac{\nu_{max}}{f_{\nu_{max}} \, \nu_{max,\odot}}\right)^3 
    \left(\frac{\Delta\nu}{f_{\Delta\nu} \, \Delta\nu_\odot}\right)^{-4} 
    \left(\frac{T_{\rm eff}}{T_{\rm eff,\odot}}\right)^{3/2},
    \label{eq8}
\end{equation}

\begin{equation}
\left(\frac{M}{M_\odot}\right) \approx 
\left(\frac{\Delta\nu}{f_{\Delta\nu} \, \Delta\nu_\odot}\right)^2 
\left(\frac{L}{L_\odot}\right)^{3/2}
\left(\frac{T_{\rm eff}}{T_{\rm eff,\odot}}\right)^{-6},
\label{eq9}
\end{equation}

\begin{equation}
\left(\frac{M}{M_\odot}\right) \approx 
\left(\frac{\nu_{max}}{f_{\nu_{max}} \, \nu_{max,\odot}}\right) 
\left(\frac{L}{L_\odot}\right) 
\left(\frac{T_{\rm eff}}{T_{\rm eff,\odot}}\right)^{-7/2},
\label{eq10}
\end{equation}

\begin{equation}
    \left(\frac{M}{M_\odot}\right) \approx 
    \left(\frac{\nu_{max}}{f_{\nu_{max}} \, \nu_{max,\odot}}\right)^{12/5} 
    \left(\frac{\Delta\nu}{f_{\Delta\nu} \, \Delta\nu_\odot}\right)^{-14/5}
    \left(\frac{L}{L_\odot}\right)^{3/10}.
    \label{eq11}
\end{equation}

For clarity, we rename the asteroseismic masses obtained from Eqs.~\ref{eq8} to~\ref{eq11} as $M_1,~M_2~,M_3,~M_4$ respectively. We adopted solar reference values of $\nu_{max,\odot}=3090\pm30~\mu$Hz, $\Delta\nu_\odot=135.1\pm0.1~\mu$Hz~\citep{2011ApJ...743..143H}, and $T_{\rm eff,\odot}=5772\pm0.8$~K~\citep{2016AJ....152...41P}. Here, $f_{\nu_{max}},~f_{\Delta\nu}$ are correction factors for $\nu_{max},~\Delta\nu$ respectively to account for any deviation from the scaling relation~\citep{2011ApJ...743..161W,2012ASSP...26.....M,2016ApJ...822...15S}. We set $f_{\nu_{max}}=1$~\citep{2018ApJS..236...42Y} and calculate $f_{\Delta\nu}$ using ASFGRID~\citep{2016ascl.soft03009S}, which estimates the correction factor by interpolation in grids of models for $-3\leq~$[Fe/H]$~\leq0.4$ and $0.6\leq\rm{M/M}_\odot\leq5.5$. Because $f_{\Delta\nu}$ depends on the evolutionary state of the star,  we need to know the evolutionary phase of the Ba stars before applying it. We use the $\Delta P$ values derived in Sect.~\ref{evs} to classify the stars into RGB or CHeB phase while correcting for $\Delta\nu$. For the Ba giants without $\Delta P$ measurements, we take the average value of $f_{\Delta\nu}$ obtained by considering that the star is in the RGB as well as the CHeB stage and assign a $1\sigma$ uncertainty of 0.02 to $f_{\Delta\nu}$ to account for its variability arising from different model assumptions in stellar models~\citep{2025A&A...698A.111V}. Likewise, Ba dwarfs may occupy either the MS or SG branch of evolution. Similarly, we have adopted the average value of $f_{\Delta\nu}$ obtained by considering the star in the  MS and SG phase while assuming an uncertainty of 0.02. The $f_{\Delta\nu}$ values for Ba giants and dwarfs are given in Table~\ref{delnu mass}.

Overall, we find that the various mass estimates are consistent with one another within the error bars (see Table~\ref{delnu mass}). We also note that for Ba giants, $M_2$ exhibits the highest uncertainty generally, compared to the other mass equations, consistent with previous studies (e.g.~\citealt{2022MNRAS.515.3184H,2023ApJ...957...18M}). The larger uncertainties in $M_2$ arise from its dependence on $\Delta\nu$, for which the required corrections introduce additional sources of error. As discussed in  Sect.~\ref{estimating delnu}, reliable determinations of $\Delta\nu$ were possible for fewer than $50\%$ of our sample.~
Additionally, the reliability of the $\Delta\nu$ corrections is limited by the unknown evolutionary states of most of the sample stars. Given these issues, and considering that the $\nu_{\mathrm{max}}$ measurements are comparatively robust (see Figure~\ref{numax compare}), we adopt $M_3$ as our primary mass estimate for the subsequent analysis and discussion.

\subsection{Modelling  Ba star systems: Description of stellar models}
\label{model description}
With reliable asteroseismic masses now estimated for our Ba stars, we can construct a grid of stellar models across the measured mass range, allowing us to investigate the behaviour of $s$-process abundance with respect to stellar mass in the binary mass transfer scenario.

We use the Modules for Experiments in Stellar Astrophysics\footnote{Version: r24.03.1} (MESA) stellar evolution code~\citep{2013ApJS..208....4P,2018ApJS..234...34P,2019ApJS..243...10P} to model Ba star systems. In particular, we model the accretion phase and investigate the behaviour and subsequent mixing of light and heavy elements post-accretion from its AGB companion. 
Unless otherwise stated, we use the same physical assumptions and numerical setup given below for all the models.

We have adopted the OPAL opacity tables~\citep{1996ApJ...464..943I} with a solar-scaled composition from~\cite{1998SSRv...85..161G} along with an Eddington-Grey $T-\tau$ relation for the atmospheric boundary condition~\citep{1978stat.book.....M}. Type 2 opacity accounts for variations in opacity as C and O abundances change throughout stellar evolution. This is essential during He burning and beyond to accurately model Ba giants found in He-core burning phases.
We use a custom nuclear network with 75 species based on the MESA agb.net network. The agb.net nuclear network includes isotopes from $^1$H to $^{22}$Ne, along with neutrons. To track the evolution of light as well as heavy elements of interest, we have added the major isotopes from $^{6}$Li to $^{209}$Bi, comprising 37 heavy isotopes beyond iron, to agb.net. We included 32 reactions to calculate the changes in composition, all involving the light elements. These reactions include the pp-chains, triple--$\alpha$ reaction, CNO cycles and $\alpha$-capture reactions (for example, $^{3}$He($\alpha, \gamma)^{7}$Be, $^{12}$C$(\alpha,\gamma)^{16}$O etc.). We do not include any reactions involving heavy elements in the nuclear network, as nucleosynthesis involving these elements is not expected prior to the AGB phase. The default mixing-length parameter, $\alpha_{\rm MLT}=2.0$, is adopted. During evolution into the giant phase, we used the Reimers mass loss with a scaling factor $\eta=0.3$~\citep{1975psae.book..229R}.

In this work, we focus on the mixing of $s$-process rich material accreted from the AGB companion onto the envelopes of Ba stars of different masses, irrespective of the mass transfer mechanism. Keeping the mass accreted ($M_{\rm acc}$) as a free parameter, we accrete different amounts of AGB material from available AGB models onto the progenitor Ba star while invoking stellar mixing mechanisms.
To produce our models, we follow~\cite{2021MNRAS.505.5554S}. We evolve the primary as a single star up to an age at which the AGB companion enters the TP-AGB phase. This age is adopted from~\cite{2014MNRAS.445..347K,2016ApJ...825...26K} and~\cite{2016ApJ...823..102C}. At this age, we started accreting AGB material onto the stellar surface at a rate of $10^{-7}~\rm{M}_\odot$yr$^{-1}$, until the current mass of the Ba star was reached. This mass accretion rate falls within the typical mass-loss rate observed in the TP-AGB phase ($10^{-8}-10^{-5}~\rm{M}_\odot$yr$^{-1}$;~\citealt{2017MNRAS.465..403G,2010A&A...523A..18D}). This also implies that the mass accretion timescale is comparable to the AGB lifetime ($\sim1-30$~Myr;~\citealt{2016ApJ...825...26K}), which is much shorter than the MS lifetime. For example, considering a constant mass loss rate of $10^{-7}~\rm{M}_\odot$yr$^{-1}$, a mass accretion of $0.1~\rm{M}_\odot$ corresponds to an accretion timescale of 1 Myr.

To constrain the combination of masses of Ba stars and their respective AGB companions, we adopt a typical mass ratio, $q=M_{\rm AGB}/M_{\rm Ba}$ of 1.3, consistent with the studies of~\cite{2019A&A...626A.127J}, and  Sarmah et al. (in prep.).
 ~\cite{2019A&A...626A.127J} reported that mild Ba stars have a $q$ value of $1-1.4$, whereas strong Ba stars have $q$ values slightly larger than 1.4. Since some of the lowest-mass Ba stars with the highest [s/Fe] in our sample exhibit the lowest metallicities, we adopt $Z\approx0.003$ for $M_{\rm Ba}<1.5~\rm{M}_\odot$, while for $M_{\rm Ba}>1.5~\rm{M}_\odot$, we use $Z\approx0.007$, which is typical of Ba stars in this sample (see Table~\ref{gba dba param} and Figure~\ref{s_feh}). Given that the available AGB models do not provide models for each required mass to exactly match $q=1.3$, we select the AGB models that provide the closest mass ratio to this value. The details of the masses of Ba stars and their companion AGB stars, together with their metallicity and [s/Fe], considered in the models are given in Tables~\ref{table:2} and~\ref{table:A2}.

\begin{table}
\caption{Mass ($M_{\rm AGB})$, metallicity (Z), and [s/Fe] of the Monash AGB models and the final mass of Ba star $(M_{\rm Ba})$ used in the MESA models}            
\label{table:2}    
\centering                        
\begin{tabular}{c c c c}      
\hline\hline        
$M_{\rm Ba} (\rm{M}_\odot)$ & $M_{\rm AGB} (\rm{M}_\odot)$ & Z& [s/Fe]   \\    
\hline                   
1 & 1.5 & 0.0028 & $1.52\pm 0.32$\\
1.5 & 1.9 & 0.007 & $1.32\pm0.41$\\
2 & 2.5 & 0.007 & $1.63 \pm0.41$\\
2.5 & 3.25 & 0.007 & $1.46 \pm 0.41$\\
3 & 4 & 0.007 & $1.38 \pm 0.41$\\
3.5 & 4.5 & 0.007 &  $0.89\pm0.41$\\
\hline
\end{tabular}
\end{table}

We construct our model Ba stars by accreting a known amount of mass: $0.05,0.1,0.3,0.5~\rm{M}_\odot$, until the final mass of $M_{\rm Ba}$ (Tables~\ref{table:2} and~\ref{table:A2}) is reached. The composition of the accreted AGB material is taken from Monash models~\citep{2016ApJ...825...26K,2018MNRAS.477..421K} that provide stellar yields and surface abundances of AGBs from masses $1-8~\rm{M}_\odot$ and $Z=0.0028-0.03$. For comparison, we also consider the AGB yields of the FRUITY (FUll-Network Repository of Updated Isotopic Tables $\&$ Yields) model available via the FRUITY database\footnote{\url{http://fruity.oa-abruzzo.inaf.it/}}. As shown in Figure~\ref{fruity_Monash}, the FRUITY yields are generally not as enriched as the Monash yields for stars of similar mass and metallicity, particularly for heavy elements. This is due to differences in model assumptions, such as the adopted mass-loss rate, which leads to different numbers of TDUs in the models. However, the light elements yielded by the two models agree with each other. We discuss the details of our stellar models with AGB composition from FRUITY in Appendices~\ref{appendix fruity mesa} and~\ref{appendixB}.

Since the accreted material has a higher mean molecular weight than the material in the envelope of the star, it is likely that a secular instability sets in, giving rise to thermohaline (TH) mixing~\citep{1972ApJ...172..165U,1980A&A....91..175K, 2021MNRAS.505.5554S,2008MNRAS.389.1828S}. TH mixing occurs on a thermal timescale, mixing the accreted AGB material into the stellar envelope until the molecular weight difference has dissipated. MESA models TH mixing using a diffusive approximation, characterised by a dimensionless parameter, $\alpha_{\rm thm}$ related to the aspect ratio of blobs arising from this instability (see~\citealt{2013ApJS..208....4P}). In this work, we have considered $\alpha_{\rm thm}=100$, which has been used to reproduce abundance patterns and Li depletion in RGB stars~\citep{2020NatAs...4.1059K}. As the range of accepted $\alpha_{\rm thm}$ is $1\leq\alpha_{\rm thm}\leq667$, we performed multiple tests to assess the influence of varying $\alpha_{\rm thm}$ on the evolution of $s$-process abundances of the modelled Ba star. We find that it has minimal effect on the final $s$-process abundances.

The models are evolved from the pre-MS to near the tip of the RGB phase. This incorporates the FDU during the RGB phase, which represents the point at which the convective envelope reaches its maximum depth into the stellar interior, leading to the maximum dilution of the accreted material. 

\section{Results}
\subsection{Evolutionary phase of Ba giants}
\begin{figure}
    \centering
    \includegraphics[width=\hsize]{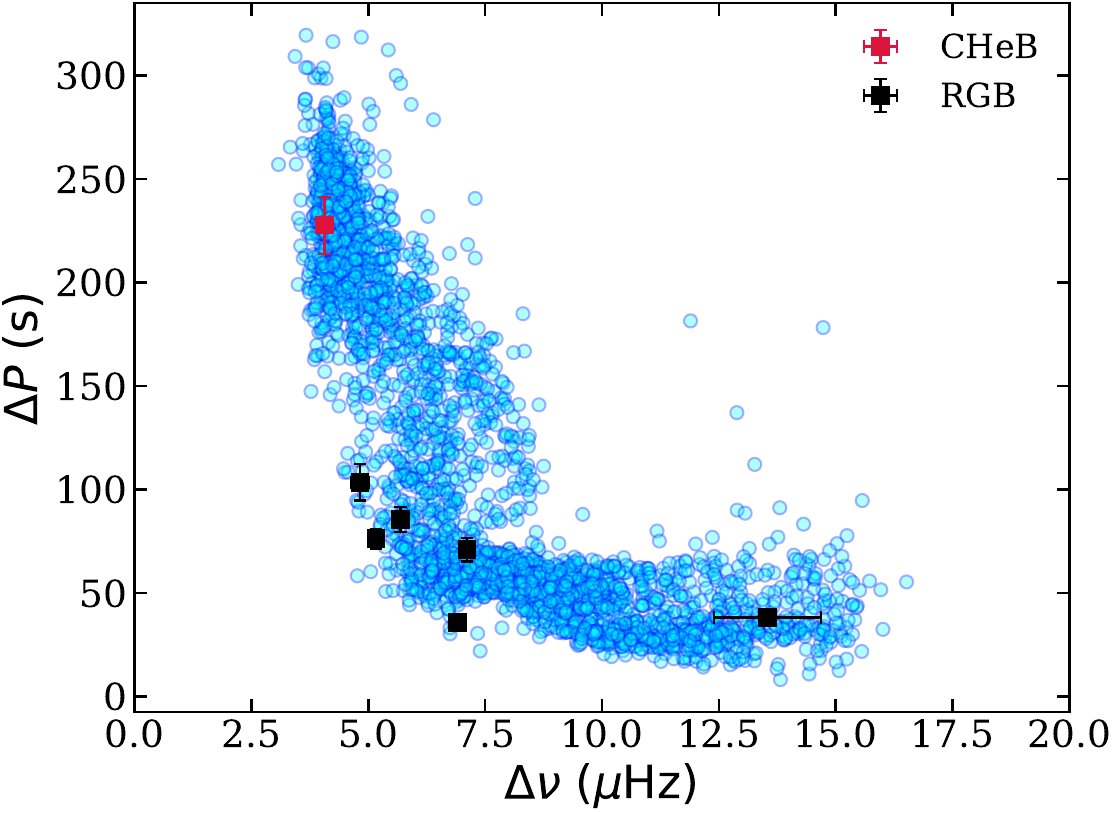}
    \caption{$\Delta P$ vs $\Delta\nu$ for Ba giants. The red square corresponds to the Ba star in the CHeB phase, and the black squares correspond to Ba stars in the RGB phase.  The background sample in blue is taken from~\cite{2013ApJ...765L..41S}.}
    \label{delp-delnu}
\end{figure}

Figure~\ref{delp-delnu} illustrates the distribution of Ba giants in $\Delta P-\Delta \nu$ plane overplotted on the asteroseismic sample of~\cite{2013ApJ...765L..41S}. We could determine $\Delta P$ confidently for seven Ba giants (see Table~\ref{delnu mass} and Figure~\ref{delp-delnu}), out of which six are in the RGB phase ($\Delta P<150~$s) and only one star is currently burning He in its core ($\Delta P > 150$~s). This shows that Ba stars are distributed across multiple evolutionary stages of the HR diagram, including the MS, SG branch, RGB, and CHeB phases (also see Figure~\ref{hrd_numax_tracks}).

\subsection{Behaviour of $s$-process abundances with mass}
\label{s observed}
\begin{figure*}
    \centering
    \includegraphics[width=\hsize]{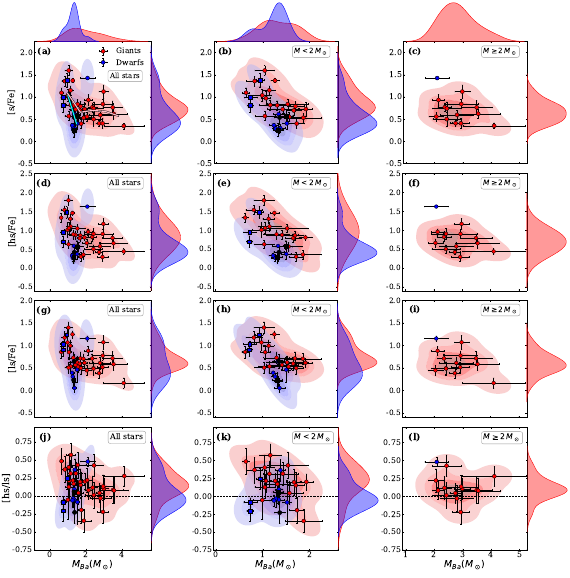}
    \caption{Behaviour of $s$-process abundances of Ba giants (red 
    circles) and Ba dwarfs (blue circles) vs their seismic mass. Here, $M_{\rm Ba}$ is the asteroseismic mass $M_3$ of Ba stars. The blue squares represent Ba 
    dwarfs with marginal detections, and the black symbols denote stars with 
    [s/Fe] estimated from fewer than three elements. \textit{Panels (a)--(c)}: [s/Fe] 
    vs $M_{\rm Ba}$ for the full sample, low-mass, and intermediate-mass 
    regimes. The mean [s/Fe] for each mass bin (bin size of $0.5~\rm{M}_\odot$ for Ba dwarfs and $0.7~\rm{M}_\odot$ for Ba giants) is shown by the cyan line for Ba dwarfs and by the pink line for Ba giants in panel (a). \textit{Panels (d)--(f)}: Same as (a)--(c), but for 
    [hs/Fe]. \textit{Panels (g)--(i)}: Same for [ls/Fe]. \textit{Panels (j)--(l)}: 
    Same for [hs/ls]. The contours and KDEs are shown 
    for giants in red and for dwarfs in blue.}
    \label{sfe_hsfe_obs}
\end{figure*}

The [s/Fe] ratio provides a measure of the average enrichment of the key $s$-process elements, including Sr, Y, Zr, Ba, La, Ce, and Nd~\citep{2018MNRAS.474.2129K,2023AJ....165..154G}. We have listed the references for the spectroscopic parameters and abundances for our sample stars in Table~\ref{gba dba param}. Figure~\ref{sfe_hsfe_obs}a shows the [s/Fe] ratio against $M_{\rm Ba}$, which is the asteroseismic mass $M_3$  for our sample.
We observe an overall anti-correlation between the [s/Fe] ratio and the seismic mass of Ba stars. There is also an offset of about 0.3~dex between the peak [s/Fe] values of Ba giants (peak at $\sim 0.7$~dex) and Ba dwarfs (peak at $\sim0.4$~dex; see the  kernel density estimations, KDEs of Figure~\ref{sfe_hsfe_obs}a).

To further investigate the dependence of $s$-process enrichment on stellar mass, we divided our entire sample into two groups: low-mass ($< 2~\rm{M}_\odot$) and intermediate-mass stars ($\geq2~\rm{M}_\odot$). For the low-mass group, we calculate Pearson correlation coefficients ($r$) of $-0.65$ for Ba giants and $-0.70$ for Ba dwarfs (Figure~\ref{sfe_hsfe_obs}b), indicating a strong negative correlation between mass and [s/Fe]. In contrast, the intermediate-mass group shows a large scatter and a weak anti-correlation ($r=-0.25$; Figure~\ref{sfe_hsfe_obs}c). It is also interesting to note that there is a wide range of [s/Fe] for a single mass, especially in the low-mass regime. To explore further, we divided the Ba dwarfs into mass bins of $0.5~\rm{M}_\odot$, and giants into mass bins of $0.7~\rm{M}_\odot$ and computed the average [s/Fe] in each bin. This is shown by the cyan and pink lines in Figure~\ref{sfe_hsfe_obs}a. A sharp decline in [s/Fe] with mass is clear in the low-mass regime, while no significant gradient is observed for $M_{\rm Ba}\geq2~\rm{M}_\odot$.

The [hs/ls] ratio is an excellent probe of neutron exposure for $s$-process nucleosynthesis in AGB stars. When the neutron exposure is sufficiently high to favour the production of second peak $hs$-elements relative to the first peak $ls$-elements, a positive [hs/ls] is expected. For our analysis, we have considered [hs/Fe] as the mean abundance of Ba, La, Ce, Nd, and [ls/Fe] as the mean abundance of Sr, Y, and Zr. We observe an anti-correlation between [hs/Fe] with mass for low-mass giants and dwarfs in Figures~\ref{sfe_hsfe_obs}d and \ref{sfe_hsfe_obs}e. Similar to [s/Fe], there is an offset between the peak value of [hs/Fe] for Ba giants and dwarfs. The [ls/Fe] ratio (Figures~\ref{sfe_hsfe_obs}g and~\ref{sfe_hsfe_obs}h) also shows an overall negative correlation with mass. From the KDEs in Figure~\ref{sfe_hsfe_obs}g, it can be seen that [ls/Fe] peaks at $\sim0.6$ for Ba giants, and $~\sim0.4$~dex for Ba dwarfs. 
In the intermediate-mass regime, there is no significant trend, just a scatter in [hs/Fe] and [ls/Fe] with mass.

Since the ratio [hs/ls] is independent of the metallicity of the Ba star, it serves as an excellent measure of the abundance patterns in the companion AGB star~\citep{2001ApJ...557..802B}. From Figure~\ref{sfe_hsfe_obs}j, we observe a mean [hs/ls] of $0.16\pm0.20$ for Ba giants, and $0.04\pm0.21$ for Ba dwarfs. The positive [hs/ls] ratio for most of our sample stars suggests the presence of a low-mass (typically $1-4~\rm{M}_\odot$) AGB companion. This is because the operation of the $^{13}$C($\alpha,n)^{16}$O as the main-neutron source leads to high neutron flux and produces [hs/ls]$~>0$ regardless of the metallicity and size of the $^{13}$C pocket~\citep{2014PASA...31...30K}.

We observe the expected anti-correlation of [s/Fe], [hs/Fe], and [ls/Fe] with metallicity of the sample stars in Figure~\ref{s_feh}. This arises because $s$-process efficiency is higher at lower metallicity (\textit{i.e.}, neutron-to-seed ratio increases as metallicity decreases), leading to stronger $s$-process enhancements at lower $[{\rm Fe/H}]$~\citep{2001ApJ...557..802B,2018A&A...620A.146C,2016MNRAS.459.4299D}. We notice that the stars with the strongest $s$-process enhancement ($>1.3$~dex) in our sample have [Fe/H]$~\leq-0.5~$dex, whereas most of the stars with [s/Fe]$~<1$~dex have [Fe/H]~$\approx-0.15\pm0.20$~dex. In contrast, the [hs/ls] ratio does not show any significant trend with changing metallicity, again implying that this ratio exclusively depends on the primary neutron source reaction rather than the metallicity of the AGB companion. 

\begin{figure}
    \centering
    \begin{subfigure}{0.9\columnwidth}
        \includegraphics[width=\hsize]{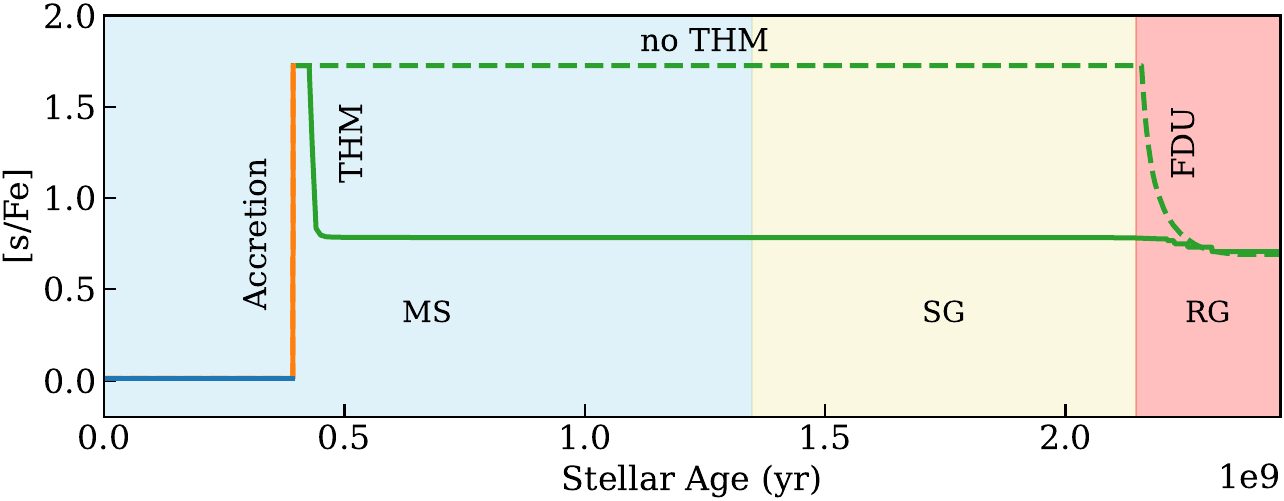}
        \caption{}
        \label{sfe_age}
    \end{subfigure}\\
    
    \begin{subfigure}{0.9\columnwidth}
        \includegraphics[width=\hsize]{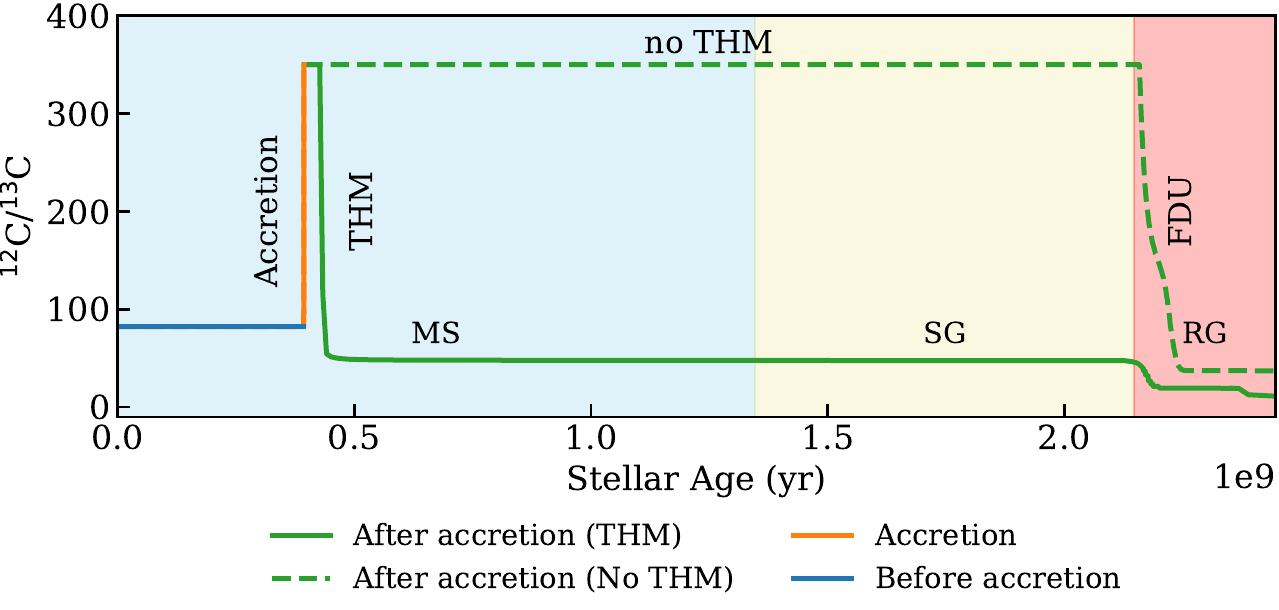}
        \caption{}
        \label{12c_13c_age}
    \end{subfigure}\\

    \begin{subfigure}{0.9\columnwidth}
        \includegraphics[width=\hsize]{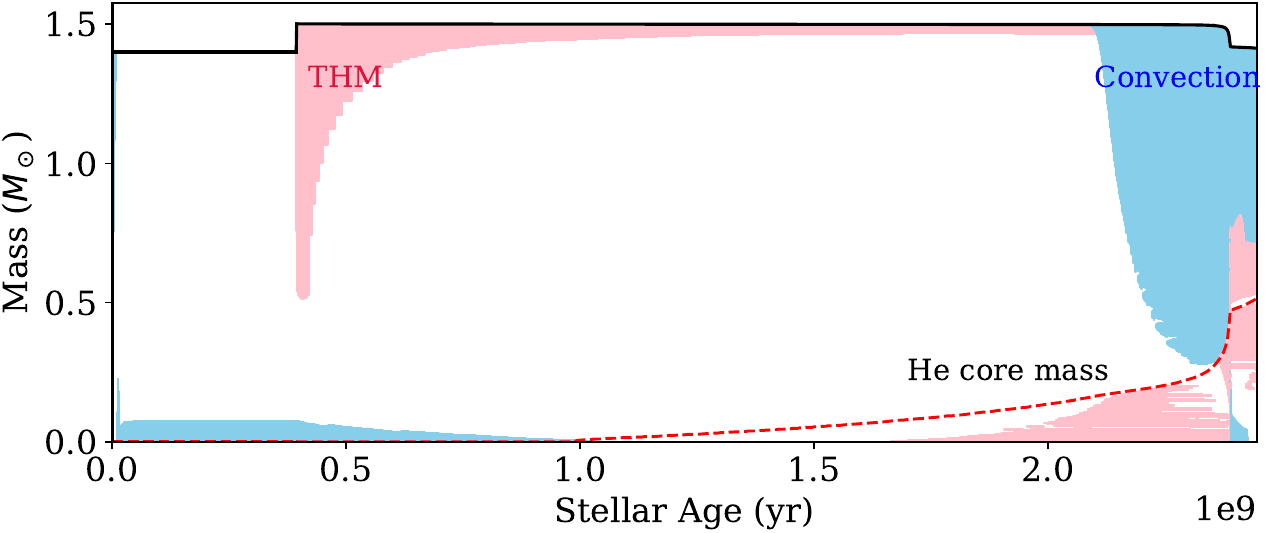}
        \caption{}
        \label{Kipp}
    \end{subfigure}\\

    \caption{\textit{Panel (a)}: Evolution of [s/Fe] with stellar age for a model with an initial mass of $1.4~\rm{M}_\odot$, and $ Z=0.007$, accreting $0.1~\rm{M}_\odot$ from a $3~\rm{M}_\odot$ AGB star. The solid blue, orange, and green lines represent the phases before, during, and after accretion in the model that considers TH mixing. The dashed lines represent the standard model without TH mixing. The light blue, yellow, and red shaded regions correspond to the MS, SG, and RG phases, respectively. \textit{Panel (b)}: Same but for $^{12}$C/$^{13}$C. Panels (a) and (b) highlight the effect of TH mixing on the C isotopic ratios, whereas the $s$-process elements are unaffected by it. \textit{Panel (c)}: Kippenhahn diagram for the same model showing regions of TH mixing in pink and convection in blue. The mass of the He-core is shown with the dotted red line.}
    \label{sfe,C_age}
\end{figure}

\subsection{Model outcomes: Effect of TH mixing on light and heavy elements}
\label{TH mix}
To investigate the observed trend of $s$-process abundances in Ba stars from the theoretical point of view, we constructed stellar models (as described in Sect.~\ref{model description}), where we found that incorporating TH mixing in our models plays a significant role in shaping the surface abundances post mass accretion.

Figure~\ref{sfe,C_age} shows the evolution of [s/Fe] and $^{12}$C/$^{13}$C for a Ba star of initial mass $1.4~\rm{M}_\odot$, and $~Z=0.007$ that has accreted $0.1~\rm{M}_\odot$ from a $3~\rm{M}_\odot$ AGB companion and has been evolved till the end of the CHeB phase. The accretion process occurs on an AGB timescale of a few million years ($<<$~MS lifetime of $\sim10^9$~years) and thus appears as a thin vertical line in that plot. From Figure~\ref{sfe,C_age} it is evident that the majority of the dilution of heavy and light elements occurs (through TH mixing) shortly after mass accretion, as the TH mixing also occurs on a much shorter timescale ($\sim 10^6$ years) than the MS lifetime. The next dilution event is the FDU, where the [s/Fe] further reduces by $~\sim0.1$~dex. Between the dilution events caused by TH mixing and FDU, the s-process abundances remain largely unchanged, preserving the nucleosynthesis signatures acquired during the accretion process. Moreover, no significant change in the abundances of $ s$-process elements is observed after FDU until the end of the CHeB phase. 

It is noteworthy that TH mixing has a negligible impact on the surface abundances of heavy elements during the giant phase. As shown in Figure~\ref{sfe_age}, the final post-FDU abundance of $s$-process elements remains essentially unchanged regardless of whether or not additional mixing processes are included in the stellar models. This is because TH mixing only redistributes the incoming $s$-process-rich material within the envelope of the Ba star, without creating or destroying it.
Therefore, the post-FDU $s$-process abundances are determined primarily by the total accreted mass and the depth of the FDU, rather than by the details of pre-FDU TH mixing.

On the other hand, the abundances of light elements and isotopic ratios after going through TH mixing are significantly affected by the FDU. For example, in Figure~\ref{12c_13c_age} we note a $^{12}$C/$^{13}$C ratio of $\sim50$ in the Ba dwarf phase, which reduces down to $\sim20$ as the star ascends the RGB and undergoes FDU.
The $^{12}$C/$^{13}$C ratio gets significantly affected by TH mixing as evident from the models with and without TH mixing in Figure~\ref{12c_13c_age}. This is because TH mixing can reach layers in the stellar interior that contain CN-processed material. It can facilitate the transport of material between the envelope and regions near the H-burning region where CN-cycle processing converts $^{12}$C to $^{13}$C, thereby changing the surface C isotopic ratio. We can clearly observe the effect of TH mixing on the $^{12}$C/$^{13}$C ratio in the dwarf phase, as it reduces the ratio to $\sim50$ in comparison to the standard model (without including TH mixing, shown by the dotted line in Figure~\ref{sfe,C_age}), which predicts a much larger value of $\sim350$. In the RGB phase, this ratio is $\sim40$ for the standard model. But for the model including TH mixing, it is considerably lower ($^{12}$C/$^{13}$C $\sim 20$). As shown in Figure~\ref{Kipp}, a subsequent TH mixing episode towards the end of the model (CHeB phase) further lowers the $^{12}$C/$^{13}$C to $~\sim11$.

\subsection{\texorpdfstring{Comparison of model $s$-process abundances with observations (Monash AGB yields)}{Comparison of model s-process abundances with observations (Monash AGB yields)}}
\label{sfe monash}
\begin{figure*}
    \centering
    \includegraphics[width=\hsize]{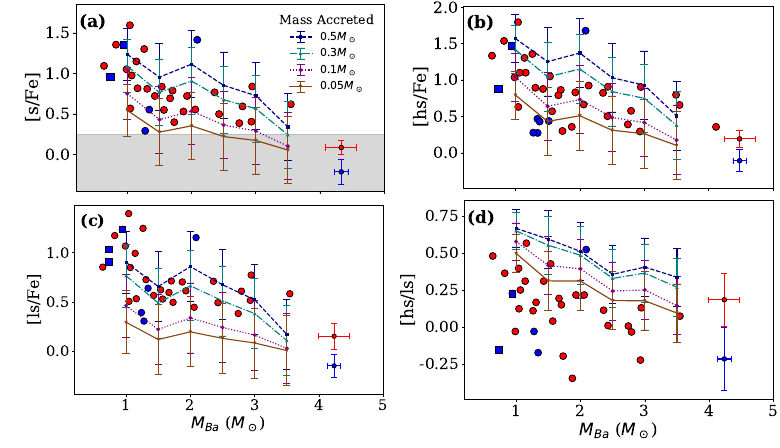}
    \caption{Comparison of $s$-process abundance ratios from MESA models using accreted AGB composition from Monash AGB yields with observed values for Ba giants (filled red circles), Ba dwarfs with clear detections (filled blue circles), and Ba dwarfs with marginal detections (filled blue squares). The representative error bars (median errors) for Ba giants and dwarfs are shown in red and blue, respectively, in the bottom right corner of each panel. The AGB donors are $\sim1.3\times M_{\rm Ba}$ as given in Table~\ref{table:2} and all models include TH mixing. \textit{Panels (a), (b), (c), and (d}): [s/Fe], [hs/Fe], [ls/Fe], and [hs/ls] vs the final mass of the Ba star after accretion ($M_{\rm Ba}$). The grey shaded region in panel (a) denotes [s/fe]$~\leq0.25~$dex. The dashed navy line, dot-dashed olive line, dotted purple line, and solid brown line correspond to an accreted mass of $0.5,0.3,0.1,0.05~\rm{M}_\odot$~respectively.}
   \label{mesa}
\end{figure*}

In Figure~\ref{mesa} we compare the observed $s$-process abundances with model predictions from MESA using AGB composition from Monash AGB yields. For consistency, we recomputed [s/Fe] in a uniform manner as the mean of Sr, Y, Zr, La, Ce, and Nd and only considered those stars for which all six elements are available. This recalculation does not affect our results, as the recalculated [s/Fe] values agree well with the initially calculated [s/Fe] values, which contain a non-uniform number of elements. Consequently, [hs/Fe] is taken as the mean of La, Ce, Nd, and [ls/Fe] is taken as the mean of Sr, Y, and Zr. The model error bars shown in Figure~\ref{mesa} reflect uncertainties in the Monash AGB yields and are propagated into the final MESA model abundances. The error bars associated with the AGB yields arise from the dispersion of $s$-process elements for stars of the same mass and metallicity but with different mixing parameters. For each mass model at a given metallicity, we calculated the standard deviation of the $s$-process abundances. Finally, the error bar for that metallicity was obtained by averaging the standard deviations across all relevant mass models. We note that this only takes into account a subset of modelling uncertainties. Since the $s$-process abundances do not vary significantly from the end of TH mixing to FDU (Figure~\ref{sfe,C_age}), we adopted our MESA model abundances near the RGB tip.

In Figure~\ref{mesa}a, it can be seen that the models reproduce the observed anti-correlation between [s/Fe] and the masses of the Ba stars. For the stars with the highest [s/Fe] $(>1$~dex), and masses around $\sim1~\rm{M}_\odot$, the models predict a mass accretion of $>0.3~\rm{M}_\odot$ (from a companion AGB star of $1.5~\rm{M}_\odot$). Such scenarios would require the Ba star to have an initial mass of $\sim 0.5-0.7~\rm{M}_\odot$~or lower, and an efficient mass transfer. For $M_{\rm Ba}>1.5~\rm{M}_\odot$, the observed [s/Fe] is in the range of $0.6\pm0.2$~dex, which is satisfied by models with $0.1-0.3~\rm{M}_\odot$ mass accretion. For the most massive Ba giants in this sample  ($3-3.5~\rm{M}_\odot$) with observed [s/Fe] greater than~$0.5$~dex, a mass accretion of $0.5~\rm{M}_\odot$ is required to match the observed [s/Fe]. This is a consequence of having a massive AGB companion ($4-4.5~\rm{M}_\odot$) with inefficient $s$-process production. Thus, the progenitor Ba star needs to accrete more to become strongly enriched in $s$-process elements.

As illustrated in Table~\ref{table:2}, although the lowest mass Ba stars ($\sim1~\rm{M}_\odot$) have AGB companions ($1.5~\rm{M}_\odot$) that do not produce the highest [s/Fe], they still exhibit the largest observed [s/Fe]. This is because low-mass stars have less envelope mass in which the accreted AGB material is diluted. Conversely, higher-mass Ba stars develop more massive convective envelopes and more extensive TH mixing. This causes the accreted AGB material to be mixed into a larger envelope mass, consequently reducing the [s/Fe] ratio on the stellar surface. From Figure~\ref{mesa}a, we note that AGB models above $5~\rm{M}_\odot$ would likely not result in the formation of a Ba star, as such AGB stars fail to synthesise enough $s$-process elements to reach the Ba star definition of [s/Fe]~$> 0.25$~dex. 

In Figure~\ref{mesa}b, we find that our models can simultaneously match the observed [hs/Fe] and [s/Fe]. Consistent with the behaviour of [s/Fe], an accretion of at least $0.1-0.3~\rm{M}_\odot$ is required from its AGB companion to explain the observed [hs/Fe] in most of the lowest mass Ba stars ($\sim1~\rm{M}_\odot)$. For the bulk of the sample ($>1.5~\rm{M}_\odot$), an accretion of $0.1-0.3~\rm{M}_\odot$ is sufficient to explain the observed $hs$ abundances, except for the most massive Ba giants, which again require an accretion of at least $0.5~\rm{M}_\odot$ to match the observed [hs/Fe].

However, this agreement is not observed for [hs/ls] (Figure~\ref{mesa}d). The observed [hs/ls] is much lower than model predictions. The mean observed [hs/ls] ratios for Ba giants and dwarfs are $\sim 0.16$ and 0.04, respectively, whereas the model [hs/ls] ratios have an average value of $~\sim 0.40$. This discrepancy arises because, although the predicted [hs/Fe] values match the observations within the uncertainties, the same models underproduce [ls/Fe] compared to the observed values (Figure~\ref{mesa}c). For instance, the $0.1~\rm{M}_\odot$ mass accretion model successfully matches the observed [s/Fe] and [hs/Fe] of most of the Ba stars (Figures~\ref{mesa}a and \ref{mesa}b), but substantially underestimates [ls/Fe]. Since the ratio [hs/ls] is the difference of [hs/Fe] and [ls/Fe], the model [hs/ls] is higher than the observed ones.

\subsection{\texorpdfstring{Comparison of model $s$-process abundances with observations (FRUITY AGB yields)}{Comparison of model s-process abundances with observations (FRUITY AGB yields)}}
\label{sfe fruity}
We present the details of our stellar models using AGB composition from FRUITY AGB yields in Appendix~\ref{appendix fruity mesa}. In Figure~\ref{mesa_fruity}, we compare the resulting model predictions with the $s$-process abundances in Ba stars. The stellar models incorporating FRUITY AGB yields fail to reproduce the observed anti-correlation between $s$, $hs$, and $ls$ with the masses of Ba stars in the low-mass regime ($<2~\rm{M}_\odot$), as shown in Figure~\ref{mesa_fruity}. This is in contrast to the results obtained using Monash AGB yields, which can successfully reproduce the observed chemical trend (see Figure~\ref{mesa}).

As explained in Sect.~\ref{model description}, this discrepancy arises because of the relatively lower $s$-process yields predicted by the FRUITY models in comparison to Monash models at a given mass and metallicity. In particular, for the $1.5~\rm{M}_\odot$ AGB companion of the lowest mass Ba star ($1~\rm{M}_\odot$), FRUITY predicts a modest amount of [s/Fe]~$=0.89$~dex (Table~\ref{FRUITY model}), which is considerably lower than the Monash yield of [s/Fe]$~=1.52$~dex (Table~\ref{table:2}). This further reduces by at least $\sim0.2~$dex relative to the pure AGB composition on taking into account dilution of the accreted AGB material within the Ba star envelope. Consequently, even under the assumption of highly efficient mass transfer ($>0.5~\rm{M}_\odot$), the FRUITY AGB yields based stellar models cannot reproduce the high $s$-process enrichment ([s/Fe]~$>1.0$~dex) in the lowest mass Ba stars (with the $1.5~\rm{M}_\odot$~AGB companion). On the other hand, the flattening of [s/Fe] and [hs/Fe] at $M_{\rm Ba}>1.5~\rm{M}_\odot$ is well reproduced with $M_{\rm acc}=0.1-0.5~\rm{M}_\odot$. Considering the large error bar associated with the seismic mass of Ba stars with $M_{\rm Ba}>3~\rm{M}_\odot$, their $s$-process abundances can be satisfied with a mass accretion of at least $0.5~\rm{M}_\odot$.

The discrepancy between model and observed abundances is particularly severe for the $ls$-elements, as evident in Figure~\ref{mesa_fruity}c. The model $ls$ abundances are significantly lower than those from the Monash-based stellar models. For the majority of the Ba star sample ($>1.5~\rm{M}_\odot$), the FRUITY yields-based models require an accretion of at least $0.5~\rm{M}_\odot$ to match the observed [ls/Fe] values, whereas the Monash-based models are able to reproduce the observations with a lower accreted mass of at least $0.3~\rm{M}_\odot$.

Since the overall yields of $ls$-and  $hs$-elements are somewhat lower in FRUITY than Monash (Figure~\ref{fruity_Monash}), the resulting model values of [hs/ls] are lower with FRUITY yields. However, it still cannot satisfactorily match the observed pattern of [hs/ls] ratio with mass of the Ba stars (Figure~\ref{mesa_fruity}d).

\subsection{Light elements in Ba stars}
\label{Light_obs_model}
The light elements, such as C, N, and O, are created and destroyed by a combination of internal mixing processes and nuclear reactions within the Ba star itself. This is in contrast to the heavy $s$-process elements, which are solely inherited from the AGB companion and are not produced/destroyed in the progenitor Ba star. To investigate the behaviour of light elements, we compiled literature abundances of C, N, O, Na, Mg, and $^{12}$C/$^{13}$C for our sample stars, whenever available (Table~\ref{gba light}). 
In case multiple abundance measurements were reported for a given star, we adopted the mean value as the abundance for that star. 

In the case of light elements, we found that the uncertainties associated with FRUITY and Monash AGB abundances (as explained in Sect.~\ref{model description}) are very small and unrealistic ($<0.1~$dex). Such small error bars do not reflect the uncertainties associated with AGB modelling arising from mass-loss rate, mixing parameters, reaction rates, etc. Therefore, to account for these uncertainties and provide a more realistic comparison, we adopted error bars comparable to the average observed abundance error bars for each element in our final MESA model abundances. 

\begin{figure*}
    \centering
    \includegraphics[width=\hsize]{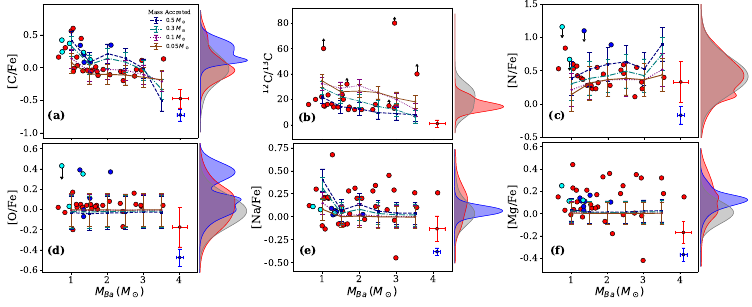}
    \caption{Comparison of light elements from MESA models with the AGB composition from Monash AGB yields with observed values for Ba giants (red) and dwarfs (blue). The Ba dwarfs with marginal detections are shown in cyan. The upward arrow corresponds to stars for which the lower limit of that abundance is available. The downward arrow corresponds to the upper limit of that abundance. The representative error bar (median error) for Ba giants and dwarfs is shown in red and blue, respectively, in the bottom right corner of each panel. The red KDE corresponds to the observed abundance of Ba giants, the blue KDE corresponds to the observed abundances of Ba dwarfs, and the grey KDE corresponds to the model abundances. The AGB donors are $\sim1.3\times M_{\rm Ba}$ as given in Table~\ref{table:2} and that all models include TH mixing. The model lines are the same as Figure~\ref{mesa}. Our stellar models, including TH mixing, can simultaneously match the $s$, $hs$ abundances and the light elements, C, N, and $^{12}C$/$^{13}$C for the majority of the Ba stars.}
    \label{mesa_light}
\end{figure*}

In Figure~\ref{mesa_light} we compare the observed light element abundances with MESA model predictions using accreted AGB composition from Monash AGB yields. From Figure~\ref{mesa_light}a, Ba dwarfs show slightly higher [C/Fe] than Ba giants, by about $0.2~$dex. This is an expected effect of the FDU, which reduces the surface $^{12}$C; since the Ba dwarfs have not undergone the FDU yet, their [C/Fe] is comparatively higher. It is evident that [C/Fe] exhibits a trend with mass of Ba star similar to that of [s/Fe] (Figure~\ref{mesa}a). This behaviour is in agreement with the fact that [C/Fe] has a direct correlation with [s/Fe]~\citep{2018MNRAS.474.2129K,2025AJ....170..259R}. This trend also indicates that Ba stars in this sample have been polluted by a low-mass AGB companion ($1-4~\rm{M}_\odot$) where $^{13}$C$(\alpha,n)^{16}$O is the primary source for neutrons for $s$-process nucleosynthesis. We further discuss this trend of C in Sect.~\ref{che trend discussion}.

\begin{figure*}
    \centering
    \includegraphics[width=\hsize]{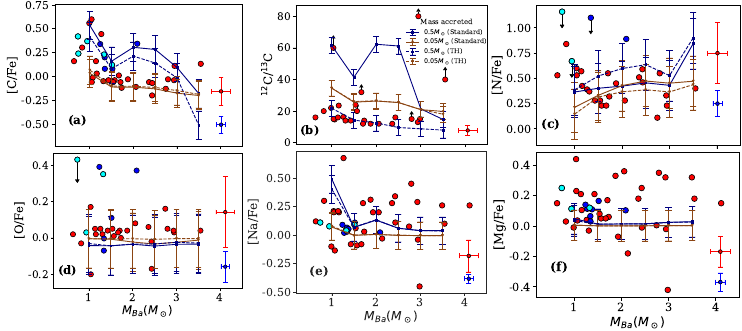}
    \caption{Comparing observed light element abundances of Ba giants (red) and dwarfs (blue) with Monash AGB yields-based MESA model values incorporating TH mixing (dashed lines) and without TH mixing (solid lines). The symbols are the same as in Figure~\ref{mesa_light}.}
    \label{mesa_light_conv_th}
\end{figure*}

In Figure~\ref{mesa_light}b, we see that Ba stars generally have a $^{12}$C/$^{13}$C ratio less than 30. For the lowest mass Ba stars (around $\sim 1~\rm{M}_\odot$), the $M_{\rm acc}=0.5~\rm{M}_\odot$ model can reproduce the observed $^{12}$C/$^{13}$C ratio, thus simultaneously matching the isotopic ratio as well as [s/Fe] (Figure~\ref{mesa}a). The $M_{\rm acc}=0.3-0.5~\rm{M}_\odot$ models can satisfactorily match the observed $^{12}$C/$^{13}$C for 85\% of our sample of Ba giants. For three Ba giants, the $^{12}$C/$^{13}$C is greater than 40, which has been attributed to a different type of TP-AGB companion \citep{2025AJ....170..259R}. Interestingly, despite their high isotopic ratio, all three stars have only a mild enhancement in C ([C/Fe]$~\leq0.7$ dex). For the two stars with $^{12}$C/$^{13}$C$~>50$ (with masses of $1.04~\rm{M}_\odot$, and $2.94~\rm{M}_\odot$), their current $\log{(L/L_\odot)}$ of 1.09$\pm$0.08, and 1.81$\pm$0.08 indicate that they are probably currently undergoing the FDU. This might be another reason for the higher $^{12}$C/$^{13}$C, as their convective envelope has not reached its deepest extent yet. We note that only lower limits on $^{12}$C/$^{13}$C are available for these stars.

As noted in Sect.~\ref{TH mix}, TH mixing post-accretion significantly reduces the $^{12}$C/$^{13}$C ratio relative to the standard model (see Figure~\ref{12c_13c_age}). We compare the model predictions of the C isotopic ratios with and without TH mixing to the observed ratios in Ba stars in Figure~\ref{mesa_light_conv_th}b. It is evident that the standard model predicts a much higher $^{12}$C/$^{13}$C than the model incorporating TH mixing ($\lesssim 30$) when a significant amount of mass is accreted ($M_{\rm acc}=0.5~\rm{M}_\odot$), thus failing to explain the observed ratios in Ba stars. In contrast, when the mass accreted is small ($M_{\rm acc}=0.05~\rm{M}_\odot$), the difference between TH mixing-incorporated and standard models is negligible. This is because the smaller amount of accreted AGB material does not produce a significant difference in the mean molecular weight to drive an efficient TH mixing (Appendix~\ref{Kepp_acc_appendix}) and the mixing does not reach layers enriched by the products of CN cycling. Overall, models including TH mixing provide a significantly better match to the observed $^{12}$C/$^{13}$C ratios in Ba stars. This arises because TH mixing not only dilutes the AGB material into the Ba star interior, but also transports material processed by the CN cycle from the stellar interior to the surface, thereby increasing the surface abundance of $^{13}$C and lowering the isotopic ratio. 

As shown in Figure~\ref{mesa_light}c, the observed [N/Fe] ratios align well with the model values for the majority of the Ba giants. Most of the Ba stars have [N/Fe] less than $0.7~$dex, consistent with pollution from low-mass AGB companions that do not experience hot bottom burning (HBB). Our models show that the N abundance increases during accretion and then decreases just after accretion due to TH mixing. It then increases to values typical of field stars right after FDU. Therefore, the observed final N abundance in Ba giants is primarily a result of the FDU.

The adopted AGB models predict fairly low Na, with [Na/Fe] $<0.5~$dex for low-mass Ba stars, and decreasing further to below 0.2 dex for intermediate-mass Ba stars. The observed [Na/Fe] of Ba giants peaks at $\sim0.2$~dex, whereas that of Ba dwarfs peaks at $\sim0.06$~dex. From the adjacent KDEs in Figure~\ref{mesa_light}e, the Na abundance in Ba dwarfs is more consistent with model values, whereas the giants show a slight overabundance of $\sim0.2~$dex, which has also been found in a previous work by~\cite{2013MNRAS.431.3338R}.
From Figure~\ref{mesa_light}d, we note that the O abundance is $~\sim 0.0~$dex for all Ba star models,  owing to negligible oxygen production in low-mass AGB nucleosynthesis. While the Ba giants match this behaviour, three of the Ba dwarfs display higher [O/Fe] than predicted. Finally, we find that most of the intermediate-mass Ba stars show a slight overabundance of [Mg/Fe] as compared to the models by $\sim0.1~$dex (Figure~\ref{mesa_light}f). This is because the AGB yield models start with an initial solar composition ([Mg/Fe]~$= 0.0$), which is lower than what is observed in field stars of similar metallicities to the Ba stars. We note that the observed [Mg/Fe] in the Ba star sample is $<0.45~$dex, which is consistent with that of field stars~\citep{2016MNRAS.459.4299D}.

We also present the model predictions using FRUITY yields for light elements in Appendix~\ref{Fruity light}. It is noteworthy that, despite discrepancies between these models in producing $s$-process elements (Sect.~\ref{sfe fruity}), they yield fairly consistent results for the light elements, such as C, N, O, Na, and Mg. However, unlike the Monash AGB models, MESA models using FRUITY AGB yields are unable to match the $s$-process abundances and light elements simultaneously.

\section{Discussion}
\label{Discussion}
The chemical peculiarities of Ba stars have been extensively studied, but a clear understanding of how their chemical anomalies relate to their stellar parameters is missing due to the lack of accurate and precise masses. Previous studies have relied on spectroscopic or HR diagram masses, which carry large uncertainties of $30-60\%$, and often much larger. The current study bridges this gap by presenting the first extensive asteroseismic study of Ba stars (31 Ba giants and 13 Ba dwarfs), deriving stellar masses with a precision of $\sim16\%$ for Ba giants and $~\sim 10\%$ for Ba dwarfs, and, for seven Ba giants, evolutionary phase through $\Delta P$ measurements. 

With the asteroseismic masses, we have investigated the Ba star mass distribution, and, by combining the asteroseismic masses with the stellar chemical abundances, also the behaviour of $s$-process and light-element abundances with stellar mass. Further, this quantitative information has enabled us to simulate the accretion and post-accretion mixing with dedicated stellar models across the measured mass range of Ba stars. Through the models, we examined the implications of additional factors such as the role of AGB yields, mass accretion, and post-accretion stellar mixing  in shaping the observed chemical trends in these stars. With this new, broad information, we revisit the origin and evolution of Ba stars. In the following subsections, we present an overview of our key results and discuss their implications.

\subsection{Mass distributions and proposed evolutionary link between Ba giants and dwarfs}
\label{int dwarfs and ev link}

Our derived asteroseismic masses (Table~\ref{gba dba param} and Figure~\ref{sfe_hsfe_obs}) place the Ba giants in the mass range of $1-4~\rm{M}_\odot$, in contrast to previous findings of $1-6~\rm{M}_\odot$ using the Kiel diagram (e.g. \citealt{2016MNRAS.459.4299D}). On comparing our seismic masses with that of~\cite{2016MNRAS.459.4299D} for a common sample of Ba giants, we found that the latter masses are on average $\sim60\%$ higher than the seismic masses, indicating that such methods may overestimate stellar masses. Ba stars of higher mass ($\gtrsim 4.5~\rm{M}_\odot$) are not expected because their AGB companions (with masses $> 4.5~\rm{M}_\odot$) are unable to synthesise enough $s$-process elements to reach the [s/Fe] value of $0.25$~dex that defines Ba stars (our Figure~\ref{mesa}a shows this trend). 

Our sample of Ba dwarfs (13 stars) and Ba giants (31 stars) shows an average mass of $1.29\pm 0.09~\rm{M}_\odot$ and $1.96\pm0.16~\rm{M}_\odot$, respectively (Table~\ref{gba dba param}).
This is in qualitative agreement with \cite{2019A&A...626A.128E} who found that the mass distribution of their sample of Ba dwarfs and giants peaks at very different masses ($\sim1~\rm{M}_\odot$ for dwarfs, and $~2~\rm{M}_\odot$~for giants). However, using our more accurate mass estimates, we identify a significant difference in the mass distribution of the giant stars.
Our distribution shows that the giants and dwarfs actually peak at a similar mass ($\sim 1.3~\rm{M}_\odot$; see the KDEs in Figure~\ref{sfe_hsfe_obs}a). The distribution of the dwarfs is similar to that of \cite{2019A&A...626A.128E}, being sharply peaked at $\sim 1.3~\rm{M}_\odot$. At lower masses, our giant star distribution is similar to that of the dwarfs, but the giants have a tail out to higher masses, as also reported by \cite{2019A&A...626A.128E}. One explanation for the disagreement between the two studies with regard to the peak of the giant mass distribution (their $\sim 2~\rm{M}_\odot$ to our $\sim 1.3~\rm{M}_\odot$) is that it is quite difficult to separate giants in mass along the RGB in the HR diagram, since different mass models/stars follow very similar paths. In particular, observationally inferred $T_{\rm eff}$ often contains substantial observational uncertainties. Meanwhile, models contain uncertainties in the use of $\alpha_{\rm MLT}$, which systematically alter stellar tracks in $T_{\rm eff}$ (among other uncertainties). Compounding this is the red clump/CHeB phase, which is difficult to distinguish from the RGB (e.g.~\citealt {2011Natur.471..608B}). Asteroseismic masses do not suffer from these uncertainties as much, and we have propagated the uncertainties from parameters like $T_{\rm eff}$ in our reported masses (Table~\ref{gba dba param} ).

Given the newly found overlap in mass distribution between the giants and dwarfs at lower masses, it is tempting to associate the low-mass Ba dwarfs with the low-mass Ba giants: they may be part of the same population. In fact, we argue that the simplest explanation for the existence of Ba giants is that they are evolved Ba dwarfs, which we refer to here as the evolutionary scenario hypothesis.

\begin{figure}
    \centering
    \includegraphics[width=\hsize]{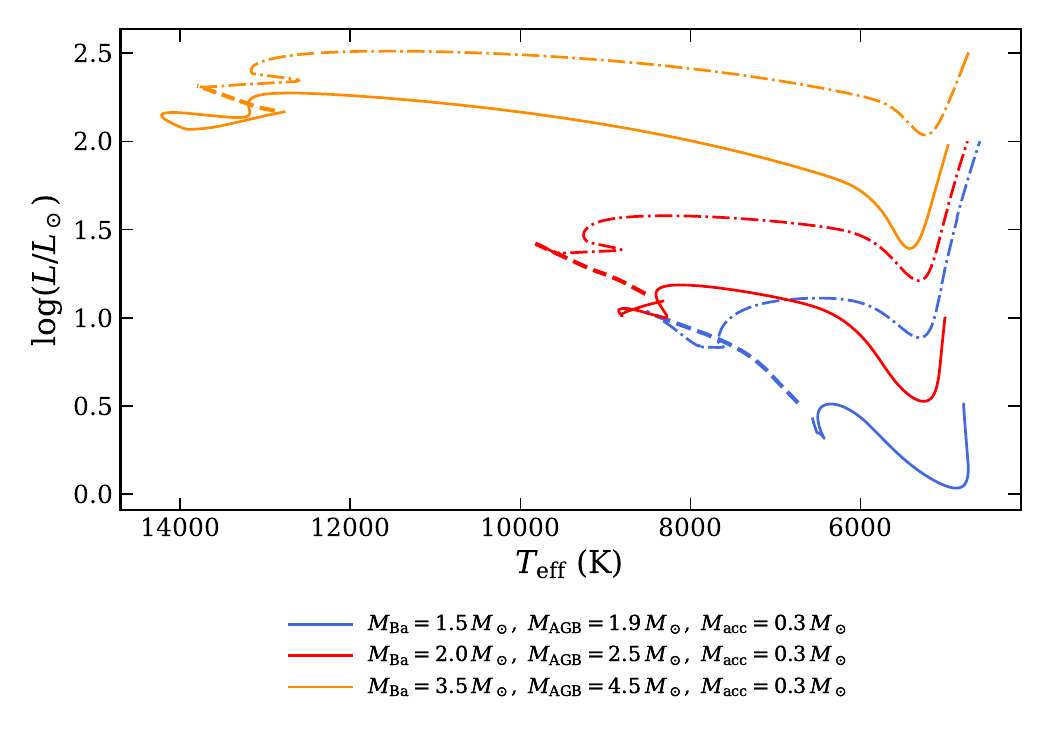}
    \caption{MESA evolutionary tracks of $1.5~\rm{M}_\odot,~2~\rm{M}_\odot, and ~3.5~\rm{M}_\odot$ (final mass) Ba stars accreting $0.3~\rm{M}_\odot$ from an AGB companion of $1.9~\rm{M}_\odot, 2.5~\rm{M}_\odot, and~4.5~\rm{M}_\odot$, respectively. The solid, thick dashed, and dot-dashed lines correspond to the phase before, during, and after accretion, respectively.}
    \label{hrd_mesa}
\end{figure}

Our models support this evolutionary hypothesis by showing that accretion is most likely to occur during the MS in binary systems with our assumed $q=1.3$ (because AGB lifetimes are relatively short; see Figure~\ref{hrd_mesa}). Varying $q$ does not alter this picture. For higher $q$ or more unequal mass systems ($q~\geq1.3$), the primary evolves significantly faster than the 
secondary, favouring accretion onto the MS companion during the AGB phase 
of the primary. For nearly equal mass systems ($q \approx 1$), the two components evolve on similar timescales, making MS accretion from an AGB donor less probable. In such 
cases, mass transfer may instead occur at earlier evolutionary stages, for instance, during the RGB phase of the donor, although the interaction depends on their orbital parameters. Therefore, systems 
with $q \approx 1$ are less likely to produce Ba stars.

The models also show that the $s$-process signatures are preserved as Ba dwarfs evolve to the giant phase (Figure~\ref{sfe_age}). Additional evidence comes from the $^{12}$C/$^{13}$C ratio in Ba giants. As explained in Sect.~\ref{Light_obs_model}, in the absence of TH mixing on the MS, the models predict substantially higher post–FDU values of $^{12}$C/$^{13}$C (Figure~\ref{mesa_light_conv_th}b), which is inconsistent with the observations of Ba giants. Only our models with MS TH mixing can reach values as low as those observed, again suggesting that accretion occurred on the MS and that the Ba dwarfs evolved into Ba giants. 

Further, the existence of a CHeB Ba star in our sample (identified through asteroseismic $\Delta P$ measurement; Table~\ref{delnu mass} and Figure~\ref{delp-delnu}), and also the recent discovery of $s$-rich RR Lyrae which have similar abundance patterns to Ba stars (\citealt{2025A&A...704A..12D}), could be explained by the further evolution of Ba giants in the evolutionary scenario.

A strong prediction of the evolutionary scenario is that the mass distribution of Ba dwarfs should match that of the Ba giants, since only minor mass-loss is expected before the upper RGB. In direct contradiction to this is the fact that only low-mass Ba dwarfs have been observed thus far, including in our sample (Figure~\ref{sfe_hsfe_obs}a; also see \citealt{2019A&A...626A.128E}).\footnote{We note that this would not be a problem if the Ba dwarfs and Ba giants were assumed to come from two separate populations.} Thus, under our evolutionary hypothesis, the whole group of intermediate-mass Ba dwarfs ($\gtrsim 2~\rm{M}_\odot$) is currently missing.

One possible reason for this deficit of intermediate-mass Ba dwarfs is observational bias. Ba dwarfs identified to date are F-, G-, and K-type stars~\citep{2018MNRAS.474.2129K, 2019A&A...626A.128E,2024AJ....167..184R}, whereas intermediate-mass dwarfs are significantly hotter and can appear as A-type stars, with surface temperatures of $>8000$~K (Figure~\ref{hrd_mesa}). At such high temperatures, it becomes difficult to isolate and detect $s$-process lines due to blending and weaker absorption lines~\citep{2018PhDT.......103H}. Moreover, A-type stars are known to be fast rotators, which broadens their spectral lines, further complicating their abundance analysis~\citep{2000IAUS..177..269N, 2018MNRAS.474.2129K}. Although spectroscopic studies of A-type stars with temperatures $>6000~$K exist in the literature~\citep{2026AN....34770060T,2023ApJ...946..110Z,2020ApJ...898...28X}, they mostly focus on Ba instead of the full set of $s$-process elements. In such stars, Ba abundances should be treated with caution as Ba is highly sensitive to non-LTE effects and also shows large overabundances at $T_{\rm eff}\gtrsim7000$~K, even in environments such as open clusters~\citep{2026AN....34770060T}. This anomaly is generally attributed to an intrinsic origin via radiative acceleration rather than an extrinsic origin. Consequently, intermediate-mass Ba dwarfs may remain undetected and misclassified among these groups of chemically peculiar stars.

A promising approach to detect the missing intermediate-mass Ba dwarfs is through extensive asteroseismic analysis with the future addition of more sector data to estimate accurate masses of such chemically peculiar dwarfs. Also needed is detailed high-resolution spectroscopic analysis together with radial-velocity monitoring of these high-temperature, intermediate-mass chemically peculiar A-type stars (e.g. $s$-rich $\delta$~Scuti stars; \citealt{2018AJ....155...45S}). Together, this data may help identify the possibly misclassified Ba dwarfs within these groups.

\subsection{Chemical trends and the role of the AGB companion}
\label{che trend discussion}
 The observed anti-correlation between the key $s$-process elemental abundances ($s,hs,ls$) with respect to mass in Figure~\ref{sfe_hsfe_obs}  emphasises the role of stellar mass in shaping the observed chemical trends. This mass trend for $s$ and [hs/ls] was also investigated by~\cite{2016MNRAS.459.4299D} using spectroscopically determined masses for Ba giants. With our significantly more accurate and precise asteroseismic masses, we find a much stronger anti-correlation ($r = -0.65$ for Ba giants, and $-0.70$ for Ba dwarfs), particularly in the low-mass regime (see Sect.~\ref{s observed}), where previous studies reported significant scatter and no clear correlation. While considering only Ba dwarfs with clear asteroseismic detection, the observed anti-correlation disappears ($r=0.15$) due to the removal of the Ba dwarfs with high [s/Fe]. However, the observed anti-correlation in Ba giants still remains strong. 
 
 One possible explanation for the comparatively low [s/Fe] at higher mass is due to the presence of a larger envelope leading to increased dilution of the incoming AGB material within the stellar interior, thereby reducing the observed surface abundances. However, by running test models, we found that merely changing the final Ba star mass while keeping the metallicity, mass of the AGB companion, and accretion mass constant only changed the [s/Fe] by about $0.3$~dex and was unable to produce the relatively strong observed slope between [s/Fe] and stellar mass. This suggests that other factors must also contribute significantly to shaping the observed $s$-process abundance trend in Ba stars.
 
Metallicity is another candidate that is anti-correlated with $s$-process element abundance, as shown by previous studies~\citep{2016MNRAS.459.4299D,2018A&A...620A.146C,2021MNRAS.507.1956R}. In the current study, we find that some of the low-mass stars with the highest [s/Fe] ($>1~$dex) are the lowest metallicity stars ([Fe/H]$<-0.5$~dex), as shown in Figure~\ref{s_feh}. This further supports the fact that $s$-process efficiency increases at lower metallicity due to a higher neutron-to-seed ratio in the companion AGB star. But it should be noted that for the bulk of the sample, with [s/Fe] less than $\sim 1~$dex, the metallicity is  $-0.15\pm0.2~$dex, suggesting that there are factors other than mass and metallicity which affect the observed chemical trend. As a sensitivity test, we ran stellar models varying only the metallicity Z from 0.007 to 0.014 while keeping the accreted mass constant across all stellar masses. We found that this magnitude of change in Z results in only a slight change in [s/Fe] of  $\sim0.05$ dex. In contrast, when we simultaneously vary the companion masses, the amount of accreted material, and the metallicity with respect to the stellar mass of the Ba star, our models successfully reproduce the observed steep slope in the $s$-process abundance as a function of stellar mass~
(see Figure~\ref{mesa}). However, as mentioned in Sects.~\ref{sfe monash} and~\ref{sfe fruity}, this is true only when we consider the composition of the accreted AGB material from Monash AGB yields. Models with FRUITY AGB yields cannot satisfactorily reproduce the observed abundances of $s$-process for the lowest mass Ba stars ($\sim1~\rm{M}_\odot$) and therefore fail to produce the observed steep slope (Figure~\ref{mesa_fruity}). This highlights the combined influence of the amount of mass accreted, AGB nucleosynthesis, together with internal mixing and stellar parameters of the Ba stars in shaping the observed chemical enrichment patterns.
 
In addition to the heavy elements, C also shows an overall anti-correlation with the mass of the Ba stars, as shown in Figure~\ref{mesa_light}a. The same models that can satisfactorily explain the observed [s/Fe], [hs/Fe] can also reproduce the observed [C/Fe] in these stars (Figure~\ref{mesa_light}a). Again, it should be noted that this is true only for our stellar models using accreted AGB composition from Monash yields (see Sect.~\ref{Light_obs_model}). This agreement is expected as C and $s$-process elements are simultaneously brought up to the surface of the AGB stars during the TDU episodes~\citep{2014PASA...31...30K,2016ApJ...825...26K}. As a result, a Ba star that has accreted a larger amount of $s$-process rich material should also show enhanced [C/Fe]. Importantly, the agreement between observed and model [C/Fe], particularly for the lowest-mass Ba stars (with an AGB companion of $1.5~\rm{M}_\odot$), provides additional validation of AGB model yields. It implies that the adopted AGB models have undergone sufficient TDU events even in low-mass AGB stars ($1.5~\rm{M}_\odot$), to match the observed C enrichment in the Ba stars.

This also indicates that the Ba stars in this sample have a companion AGB within $1-4~\rm{M}_\odot$. Higher-mass AGB stars ($>4.5~\rm{M}_\odot$) experience HBB, which prevents the formation of a $^{12}$C rich AGB by converting $^{12}$C into $^{14}$N and therefore cannot produce the observed correlation between [C/Fe] and [s/Fe]~\citep{2014PASA...31...30K}. In other words, simultaneous measurements of C and $s$-process abundances in Ba stars provide a way of constraining the masses of AGB stars undergoing TDU episodes and HBB. The presence of a low-mass AGB companion is also suggested by the low [N/Fe] ($<0.7$~dex) and positive [hs/ls] ratio in most of our Ba stars (see Sections~\ref{Light_obs_model} and~\ref{s observed}), indicating the primary source of neutrons for $s$-process in the companion AGB is $^{13}$C$(\alpha,n)^{16}$O as shown by previous studies~\citep{2018A&A...618A..32K,2024AJ....167..184R}.

We find three Ba dwarfs showing an overabundance of [O/Fe] ($>0.3~$dex; Figure~\ref{mesa_light}d) relative to model abundances ($\sim0.0$~dex). To check if these are $\alpha$-enhanced stars, we looked into the abundances of other $\alpha$-elements (Mg, Ca and Ti). We found that they behave like normal field stars in terms of these $\alpha$-elements ([Mg/Fe]$\sim0.1$~dex, [Ca/Fe], and [Ti/Fe] $<0.1~$dex), indicating that the three Ba dwarfs are not $\alpha$-enhanced. On comparing with the studies of~\cite{2026enap....2..744K} and \cite{2003MNRAS.340..304R}, which focus on Galactic chemical evolution, we found that the relatively high [O/Fe] in these three Ba stars is most likely a metallicity effect; stars at their metallicities ([Fe/H]=$-0.3$~to~$-0.5~$dex) are expected to have [O/Fe]~$\sim0.3~$dex, whereas the AGB models consider an initial O abundance of $0.0$~dex, even for sub-solar metallicities (i.e. scaled solar).

\subsection{The critical role of thermohaline mixing}
Our observations and models support the occurrence of additional mixing in Ba dwarfs soon after accretion on the MS. 
The first evidence comes from the observed abundance of $s$-process elements in Ba dwarfs. In the absence of additional mixing post-accretion, Ba dwarfs would retain almost pure AGB composition and therefore show much higher [s/Fe] than Ba giants (see Figure~\ref{sfe_age}). This is due to the fact that convective envelopes of MS stars are very shallow, so almost no dilution can occur. Instead, we observe that Ba dwarfs show similar $s$-process enrichment to Ba giants (Figure~\ref{sfe_hsfe_obs}a; with a small offset of $\sim0.3$~dex). Our models show that this can be explained by efficient TH mixing, which redistributes the accreted AGB material throughout the stellar interior. In fact, the mixing occurs over a mass similar to that of FDU (see also \citealt{2021MNRAS.505.5554S}).
With the inclusion of TH mixing, most of the dilution occurs just after accretion, on a thermal timescale, which is very short compared to the MS lifetime (Figure~\ref{sfe,C_age}). The star retains most of its surface [s/Fe] as it evolves into a giant -- FDU only reduces [s/Fe] by about $0.1$~dex (Figure~\ref{sfe_age}), indicating that FDU dilution occurs over only a small additional mass compared to TH mixing on the MS. This provides strong supporting evidence for the Ba dwarf $\rightarrow$ Ba giant evolutionary scenario proposed in Sect.~\ref{int dwarfs and ev link}. It also provides strong support for the occurrence of extra mixing through the mostly radiative interiors of low-mass MS stars, as a response to accretion.

Chemical abundance differences between Ba dwarfs and giants, based on model predictions, could provide useful tests of the evolutionary scenario. Since heavy element abundances can not be altered at the low interior temperatures in stars in this mass range, they can only act as tracers for dilution. As noted above, unfortunately, there is only a minor difference in [s/Fe] in our models between the MS and post-FDU. Thus, the abundance of $s$-process elements in evolved Ba stars can provide no substantial constraints on whether TH mixing was active or not in their dwarf phase after accretion. 

In contrast to heavy elements, the light elements (e.g. C, N), and their isotopic ratios, are more sensitive to the effects of internal mixing and provide important constraints on the presence of TH mixing in stars (e.g. \citealt{2009MNRAS.394.1051S,2021MNRAS.505.5554S}).
In Figure~\ref{mesa_light_conv_th} we compare observed Ba star abundances with model predictions with and without TH mixing, for mass accretions of $0.05~\rm{M}_\odot$ and $0.5~\rm{M}_\odot$. The abundances of C, N, and the $^{12}$C/$^{13}$C ratio are affected by TH mixing but only in the models that experience a substantial amount of mass accretion ($0.5~\rm{M}_\odot$). In contrast, the low accretion model ($0.05~\rm{M}_\odot$) shows a negligible difference between the standard and TH mixing cases. As explained in Sect.~\ref{Light_obs_model}, lower mass accretion leads to a weak mean molecular weight gradient and hence the TH mixing is inefficient (Figure~\ref{Kipp_acc}). Therefore, the abundances remain largely unchanged from the standard models. 

Considering the $0.5~\rm{M}_\odot$ accreted mass models, the addition of TH mixing decreases [C/Fe] by $\sim0.2$~dex, and increases [N/Fe] by $\sim0.15~$dex. This is expected as TH mixing reaches down to regions that have already been CN-processed (leading to gas rich in $^{14}$N and relatively poor in $^{12}$C). This material is brought up from the deep stellar interior, affecting the surface abundances. However, the magnitudes of these surface abundance differences ($0.15-0.20~$dex) are comparable to typical observational error bars, which makes it difficult to use [C/Fe] or [N/Fe] as probes for TH mixing in stellar interiors post-accretion.

On the other hand, the $^{12}$C/$^{13}$C ratio does offer some diagnostic power for TH mixing, resulting in two observational tests: one for Ba dwarfs, and one for Ba giants.
As evident from Figure~\ref{12c_13c_age} (initial mass $=1.4~\rm{M}_\odot$ model), the addition of TH mixing during the MS significantly reduces the C isotopic ratio from around 350 (essentially the AGB material value) in the standard model, to about 50. This is a very large difference that should be easily detectable. Unfortunately, none of our Ba dwarfs has $^{12}$C/$^{13}$C reported as yet. This is an important abundance ratio to measure in future work, which we are pursuing.

As the stars become giants, they undergo FDU. In the TH mixing model, this reduces the ratio further to about 20, whereas the final abundance in our standard model is significantly higher, around 40. Although a smaller difference than for the dwarf case, this should still be easily detectable. Fortunately the majority of our Ba giants have $^{12}$C/$^{13}$C measured, as shown in Figures~\ref{mesa_light}b and \ref{mesa_light_conv_th}b. As mentioned earlier, the general picture is that Ba giants have low ratios ($\sim 10-20$). In Figure~\ref{mesa_light_conv_th}b, we also overplot some representative models with and without TH mixing. Focusing on the $0.5~\rm{M}_\odot$ accreted mass models, we see that only the models with TH mixing included can reach the observed low values. The no-TH mixing models can only come close when the mass accretion is low, where they match the TH-mixing models. Although the TH mixing models can best bracket the observations (especially considering the model uncertainties), there appears to be some residual disagreement, in that the observations tend to be at the extreme end of the accretion mass ($0.5~\rm{M}_\odot$; lower $^{12}$C/$^{13}$C ratios), whereas this is not true for other abundances. \cite{2021MNRAS.505.5554S} suggested that the bias to very low ratios may be due to some missing CN-processing in AGB models (e.g. deep mixing along the AGB; ~\citealt{Karakas2010}). Despite this small discrepancy, we conclude that the $^{12}$C/$^{13}$C ratio provides yet more support for the existence of post-accretion mixing on the MS. We note that for higher mass donor AGB stars (~$\gtrsim$ 4~M$_\odot$), the difference in ratio between the TH mixing model and the standard model becomes smaller (Figure~\ref{mesa_light_conv_th}b). This is due to the AGB model yields not having high $^{12}$C/$^{13}$C ratios in the first place. This was also pointed out by \cite{2021MNRAS.505.5554S}. However, with our new precise (asteroseismic) Ba star masses, we do not expect many donors at such high masses (assuming $q \lesssim1.5$). Moreover, as noted in Sect.~\ref{sfe monash}, Ba stars are not even expected to form from donors $\gtrsim 5~\rm{M}_\odot$, due to their low s-enrichment.

Another diagnostic for discerning whether giants underwent post-accretion TH mixing on the MS is the $^{14}$N/$^{15}$N isotopic ratio. Our models predict a substantial increase in this ratio when TH mixing is included. For example, $^{14}$N/$^{15}$N increases by a factor of about 4 as compared to a typical value of $\sim1000-2000$ in the standard model for a $2~\rm{M}_\odot$ Ba star that accreted $0.5~\rm{M}_\odot$ from its AGB companion. As yet, there have been no measurements of this ratio in Ba giants; this is also an important avenue for future spectroscopic work.

Finally, similar to the heavy elements, the post-FDU O, Na, and Mg abundances are largely unaffected by the addition of TH mixing (Figure~\ref{mesa_light_conv_th}). This is because they are modified by nuclear reactions that occur at higher temperatures (Ne-Na and Mg-Al cycles) than are available in low-mass MS stars. Thus, they cannot be used as a diagnostic for the occurrence (or lack thereof) of post-accretion additional mixing on the MS. 

\subsection{\texorpdfstring{Implications of the light $s$-process underproduction in AGB models}
{Implications of the light s-process underproduction in AGB models}}
Our models suggest that in order to explain the observed trend and abundances of $s$ and $hs$, the majority of the Ba stars should have accreted in the range $0.1-0.5~\rm{M}_\odot$ from their companion. This is in agreement with previous work (e.g.~\citealt{2025A&A...701A..52D,2021MNRAS.505.5554S}).

However, we find that our models consistently and systematically underproduce the $ls$-elements. This in turn causes the model [hs/ls] ratio to be much higher than observations (Figure~\ref{mesa}d). Recently, a similar discrepancy was noted by \cite{2024A&A...688A.164V}, who compared the observed abundances of 180 Ba stars to those of Monash and FRUITY AGB models through machine learning and found that the computed residuals indicated an underproduction of some elements just after the first peak of the $s$-process elements (particularly Nb, Mo, and Ru) in the models relative to the observations. Moreover, the $ls$-elements did not follow the expected pattern of $s$-process elements. The authors attribute this to a possible trace of the intermediate neutron-capture process ($i$-process), maybe due to a late thermal pulse, or an unidentified nucleosynthesis or mixing effect. An earlier study by \cite{2023A&A...672A.143D} came to a similar conclusion, reporting discrepancies between the Ba star observations and the theoretical predictions of nucleosynthesis in the first $s$-process peak. Further evidence of a model-observational discrepancy in $s$-process production comes from post-AGB (PAGB) stars. For example, \cite{Menon2025} found that the model yields can not reproduce the [hs/ls] ratios in the majority of PAGB stars (see their Figure~9). 

Given this convergent evidence, it appears that either (i) our AGB models are not reliable for light-$s$ elements, or (ii) Ba stars do not show a standard AGB $s$-process production pattern.
Option (ii) could come about through the fact that Ba stars are not single stars but are formed in binary systems. Moreover, these systems are relatively close (some probably close enough to have undergone Roche-lobe overflow mass transfer), making dynamical factors more prominent in their evolution. For example, tidal interactions could alter the mass-loss rate and rotation. Recent work by \cite{2025PASA...42..168O} highlights how binary evolution can alter nucleosynthetic yields. It may be that Ba stars are not the ideal tests of AGB nucleosynthesis as commonly assumed.

On the other hand, we find that most Ba stars are explained by mass accretion amounts consistent with wind mass transfer, suggesting that very close interactions may not be common. If this is the case, then the bulk of Ba stars could be expected to be good tracers of single-star AGB nucleosynthesis. Offering further support for this are the observations of PAGB stars, where the [hs/ls] distributions are similarly offset to lower values compared to the models (see our Figure~\ref{mesa}d and Figure~9 of \citealt{Menon2025}). In this case, we would need to look at factors that could alter the $ls$ production in the models. One possible explanation is rotationally induced mixing, which decreases the neutron exposure due to extra mixing of the neutron poison $^{14}$N, into the $^{13}$C pocket during the interpulse period. This can alter the [hs/ls] ratio \citep{2003ApJ...593.1056H,2004A&A...415.1089S}. Another possibility is that there are problems with the $s$-process reaction rates and resultant nucleosynthetic pathways.

Beyond model uncertainties, enhanced first peak $s$-elements can also be a sign of pollution after a very late thermal pulse, just before the companion AGB becomes a WD~\citep{1995A&A...297..727B,2023A&A...672A.143D}. During this event, some degree of H-ingestion, and possibly $i$-process, occurs, as
demonstrated by Sakurai’s object~\citep{2011ApJ...727...89H,2006PASP..118..183W}. Thus, a Ba star that was polluted during the late thermal pulse phase of its companion AGB might exhibit $ i$-process signatures. However, these complex events require investigation through detailed hydrodynamic simulations to check their feasibility~\citep{2019MNRAS.488.4258D}. Moreover, this scenario is expected to occur in only $\sim 20\%$ of AGB stars, whereas we note enhanced $ls$ in the majority of our Ba star sample (Figure~\ref{mesa}c).

Because these are Population I stars, the enhanced $ls$-elements might also partially reflect pre-enrichment of their natal molecular clouds by the ejecta of massive stars. One of the well established sites for the weak $s$-process are massive stars ($\geq10~\rm{M}_\odot$) with $^{22}$Ne$(\alpha,n)^{25}$Mg as the dominant neutron source. These are known to be the main producers of $ls$-elements from Sr to Zr~\citep{1991ApJ...371..665R,2010ApJ...710.1557P,2023ARNPS..73..315L}. Theoretical nucleosynthesis predictions have also confirmed that more than $70 \%$ of Cu, Ge, Ga, As, and Rb are produced by this weak process in our Solar System. Detailed abundance analyses of these elements in Ba stars exhibiting enhanced $ls$-elements are critical to understand the respective roles of AGB, $i$-process, and massive-star contributions in shaping their chemical anomalies.

\section{Conclusion}
\label{Conclusion}
We presented the first extensive asteroseismic analysis of a sample of Ba stars (31 Ba giants and 13 Ba dwarfs) to derive their accurate and precise stellar masses. For 7 of the Ba giants, we could also ascertain the evolutionary phase. Using the constraints on the masses of Ba stars, we then constructed stellar models of Ba stars, assuming a~$q$ value of 1.3 and accreting AGB compositions from Monash and FRUITY AGB yields. Our key findings are as follows:
\begin{itemize}
    \item With the assumed $q$ value, we found that the mass accretion process took place when the progenitor Ba star was in the MS phase for all our stellar models. Combined with the newly found overlapping mass distributions of Ba giants and dwarfs in the low-mass regime (both peaking at $1.3~\rm{M}_\odot$) and low $^{12}$C/$^{13}$C observed in Ba giants (Figure~\ref{mesa_light_conv_th}b), this implies that Ba dwarfs are the progenitors of Ba giants.
    In this scenario, the Ba dwarfs evolve through the SG, RGB, and CHeB phases (with some showing up as RR Lyrae stars with $s$-enrichment), and finally, onto the AGB.

    \item Our evolutionary scenario hypothesis predicts that the mass distribution of Ba dwarfs should match that of Ba giants.
    However, we found no intermediate-mass Ba dwarfs. We attribute this to an observational bias that is probably due to their high surface temperatures.

    \item The models using the AGB yield composition from the Monash grid simultaneously reproduced the observed abundance trends of light elements, [s/Fe] and [hs/Fe], with a mass accretion of $M_{\rm acc}=0.1-0.5~\rm{M}_\odot$, in the majority of our Ba giants and dwarfs. On the other hand, stellar models incorporating FRUITY yields did not match the observed $s$-process enrichment for the lowest-mass Ba stars ($\sim 1~\rm{M}_\odot$), and they hence did not reproduce the observed anti-correlation between [s/Fe] and mass. 
However, both sets of models failed to reproduce the observed [ls/Fe], and, consequently, [hs/ls], which is consistent with previous studies of Ba and PAGB stars. The AGB models apparently underproduce the light $s$-elements. This can be attributed to missing input physics or nucleosynthesis processes in AGB models (e.g. $ i$-process, neutron poisons, or rotation), pollution from a late thermal pulse from the companion AGB, or contamination of the natal molecular cloud by ejecta from massive stars.

    \item The observed trend between [C/Fe] and stellar mass of Ba stars strengthens the correlation between C and [s/Fe]. Combined with the fact that the majority of sample Ba stars have [N/Fe]~$<0.7~$dex and [hs/ls]~$>0$, this suggests that these stars had an AGB companion within the mass range $1-4~\rm{M}_\odot$.

    \item  The inclusion of TH mixing in our models is critical for explaining the observed abundances in Ba dwarfs and the low carbon isotopic ratios in giants. 
    Stellar models using FRUITY and Monash yields while incorporating TH mixing were able to reproduce the observed low $^{12}$C/$^{13}$C ratio in the majority of Ba giants with an accretion of $0.3-0.5~\rm{M}_\odot$ from the companion AGB. This strongly suggests the occurrence of additional mixing in the MS stars post-accretion.
\end{itemize}
Overall, the chemical trends in these stars are an outcome of the combined effects of stellar structure, metallicity, amount of accreted mass, and post-accretion mixing in the Ba star, together with the stellar parameters and nucleosynthesis in the AGB companion. Therefore, it is crucial to measure precise stellar parameters for Ba stars and their companions for forming a complete picture of their formation, evolution, and chemical anomalies. 

In the continuation of this study, we plan to estimate the masses of the WD companions, which is now facilitated through the seismic Ba star masses. We will then combine the known masses with dedicated stellar models to accurately model each Ba-star binary system individually. Further, we also plan to measure carbon isotopic ratios in Ba dwarfs through high-resolution spectroscopy. These  approaches will enable us to test the evolutionary scenario proposed in the current study and infer the mass transfer mechanism(s) that cause the observed chemical anomalies. This will ultimately constrain our understanding of binary evolution in this regime.

\begin{acknowledgements}
We thank the anonymous referee for their constructive comments, which greatly improved this manuscript. LS thanks Raghubar Singh and Anohita Mallick for helpful discussions. We thank
Evgenii Neumerzhitckii for the use of his plotting routine.
SWC acknowledges federal funding from the Australian Research Council through Discovery Projects DP190102431 and DP210101299. This research was partly supported by the use of the Nectar Research Cloud, a collaborative Australian research platform supported by the National Collaborative Research Infrastructure Strategy (NCRIS). SM acknowledges the funding from the National Natural Science Foundation of China (NSFC) under grant No. 12588202 and the National Key R\&D Program of China Nos. 2023YFE0107800 and 2024YFA1611900. 
\end{acknowledgements}

\bibliographystyle{aa}
\bibliography{bibliography} 

\begin{appendix}
\onecolumn
\section{Asteroseismic measurements}
Table~\ref{gba dba param} presents the spectroscopic parameters, derived $\nu_{max}$ (Sect.~\ref{estimating numax}) and $M_3$ (Sect.~\ref{scaling rel}) of the entire sample. In the following subsections, we discuss the estimation of $\nu_{max}$ and $\Delta\nu$ using pySYD and our $\Delta P$ measurements. We also present the asteroseismic masses obtained from the scaling relations given in Sect.~\ref{scaling rel} for the sub-sample of Ba stars having $\Delta\nu$ measurements.
\begin{table*}[!htpb]
\centering
\small
\caption{Spectroscopic parameters, $\nu_{max}$ and $M_3$ for the sample of Ba giants (\textit{top}) and dwarfs (\textit{bottom}).} 
\label{gba dba param}
\begin{tabular}{lcccrcrrc} 
\hline\hline
\multicolumn{1}{l}{Object} & Detection &[s/Fe] & [Fe/H] &\multicolumn{1}{c}{ $T_{\rm eff}$~(K)}  & Ref &\multicolumn{1}{c}{$L (L_\odot)$} & \multicolumn{1}{r}{$\nu_{max}~(\mu$Hz)}  &  $M_3~(\rm{M}_\odot)$  \\
\hline
\multicolumn{9}{c}{Ba Giants} \\
\hline
HD 32712 & Clear &$0.82\pm0.11$ & $-0.24 \pm 0.16$ & $4600 \pm 90$ &2 & $68.6 \pm 13.1$ & $23.60 \pm 0.93$ &  $1.16 \pm 0.24$  \\
HD 83548 & Clear & $0.62\pm0.11$ & $+0.03 \pm 0.14$ & $5000 \pm 100$ & 2& $155.0 \pm 29.1$ & $42.77 \pm 3.73$  & $3.55 \pm 0.77$ \\
HD 36650 & Clear &  $0.57\pm0.10$ &  $-0.28 \pm 0.13$ & $4800 \pm 100$ & 2 & $51.5 \pm 9.7$ & $32.87 \pm 0.94$  &  $1.04 \pm 0.21$  \\
HD 116713 & Clear & $1.03\pm0.15$ & $-0.12 \pm 0.13$ & $4790 \pm 50$ & 3 & $41.8 \pm 7.9$ & $38.06 \pm 1.34$  & $0.99 \pm 0.19$   \\
HD 22772 & Clear & $0.79\pm0.11$ & $-0.17 \pm 0.13$ & $4800 \pm 100$ & 2 & $55.5 \pm 3.8$ & $49.75 \pm 1.83$  & $1.70 \pm 0.18$ \\
HD 24035 & Clear & $1.36\pm0.11$ & $-0.23 \pm 0.15$ & $4700 \pm 100$ & 2 & $25.2 \pm 4.9$ & $49.21 \pm 1.08$    & $0.82 \pm 0.17$ \\
HD 30554 & Clear & $0.69\pm0.10$ & $-0.12 \pm 0.14$ & $4800 \pm 100$ & 2 & $44.3 \pm 2.7$ & $61.02 \pm 1.30$    & $1.67 \pm 0.16$  \\
HD 199394 & Clear & $0.93\pm0.13$ &$-0.02 \pm 0.12$ & $5080 \pm 100$ & 4 & $50.4 \pm 2.8$ & $83.78 \pm 3.95$  &$2.14 \pm 0.21$  \\
HD 65314 &  Clear &$0.74\pm0.15$& $-0.12 \pm 0.11$ & $5000 \pm 120$ & 5 & $55.9 \pm 10.5$ & $81.33 \pm 3.23$  &   $2.43 \pm 0.51$  \\
HD 53199 & Clear &$0.77\pm0.10$ & $-0.23 \pm 0.13$ & $5000 \pm 100$ & 2 & $52.8 \pm 3.6$ & $83.50 \pm 2.35$   & $2.36 \pm 0.24$  \\
HD 39778 &Clear & $0.84\pm0.10$ & $-0.12 \pm 0.12$ & $5000 \pm 100$ & 2 & $64.1 \pm 12.0$ & $85.85 \pm 2.75$  & $2.94 \pm 0.60$  \\
HD 160507 &Clear & $0.73\pm0.15$ & $+0.09 \pm 0.14$ & $5044 \pm 47$ & 6 &$43.5 \pm 2.1$ & $85.59 \pm 3.59$  & $1.93 \pm 0.14$  \\
HD 29685 &Clear & $0.55\pm0.10$& $-0.07 \pm 0.14$ & $4900 \pm 100$ & 2 & $23.1 \pm 2.5$ & $118.54 \pm 15.62$  & $1.57 \pm 0.29$  \\
HD 107541 &Clear & $1.60\pm0.11$ & $-0.63 \pm 0.11$ & $5000 \pm 100$ & 2 & $12.2 \pm 2.3$ & $159.29 \pm 5.93$  & $1.04 \pm 0.21$  \\
HD 16458 &Clear &  $1.37\pm0.10$& $-0.64 \pm 0.11$ & $4550 \pm 25$ & 7 &$125.1 \pm 6.7$ & $13.57 \pm 0.62$   &$1.26 \pm 0.09$ \\
HD 111315 &Clear & $0.50\pm0.15$ & $+0.04 \pm 0.09$ & $4900 \pm 100$ & 2 & $221.5 \pm 41.8$ & $19.10 \pm 0.30$  &  $2.43 \pm 0.49$\\
BD-083194 &Clear & $1.09\pm0.11$ & $-0.10 \pm 0.16$ & $4900 \pm 100$ & 2 & $58.0 \pm 2.9$ & $19.13 \pm 1.42$ &   $0.64 \pm 0.07$   \\
HD 66812 &Clear & $0.35\pm0.06$ & $-0.02 \pm 0.12$ & $4960 \pm 100$ & 1 &$305.6 \pm 57.2$ & $24.46 \pm 4.48$  & $4.11 \pm 1.12$  \\
CD-611941 &Clear & $0.84\pm0.11$ &$-0.20 \pm 0.14$ & $4800 \pm 100$ &2 & $95.9 \pm 24.7$ & $25.97 \pm 2.68$  & $1.54 \pm 0.44$  \\
HD 104979 &Clear &$0.97\pm0.15$ & $-0.26 \pm 0.10$ & $5060 \pm 100$ & 8 &$57.2 \pm 3.3$ & $36.33 \pm 1.06$ &  $1.07 \pm 0.10$  \\
HD 213084 &Clear & $0.81\pm0.10$ &$-0.09 \pm 0.15$ & $5000 \pm 100$ & 2 & $58.3 \pm 11.2$ & $42.02 \pm 2.47$  & $1.31 \pm 0.28$   \\
HD 50075 &Clear & $0.72\pm0.10$ & $-0.16 \pm 0.11$ & $4900 \pm 100$ & 2 & $58.8 \pm 3.7$ & $42.31 \pm 9.13$  & $1.43 \pm 0.34$  \\
HD 177192 & Clear &$0.53\pm0.10$ & $-0.17 \pm 0.20$ & $4700 \pm 100$ & 2 & $65.7 \pm 12.6$ & $42.95 \pm 1.73$  & $1.87 \pm 0.39$  \\
HD 20394 &Clear & $1.15\pm0.10$ & $-0.22 \pm 0.12$ & $5100 \pm 100$ & 2 & $52.6 \pm 3.1$ & $43.47 \pm 2.61$ &   $1.14 \pm 0.12$   \\
HD 101013 &Clear & $1.12\pm0.15$ &$-0.40 \pm 0.10$ & $4722 \pm 32$ & 9 & $102.4 \pm 7.8$ & $44.58 \pm 1.74$ &  $2.98 \pm 0.27$   \\
HD 62017 & Clear &$0.76\pm0.15$ & $-0.32 \pm 0.14$ & $5000 \pm 100$ & 2 & $144.3 \pm 19.3$ & $45.25 \pm 2.19$ &  $3.49 \pm 0.55$ \\
HD 90167 &Clear & $0.40\pm0.11$ &$-0.04 \pm 0.11$ & $5000 \pm 100$ & 2 & $66.5 \pm 2.7$ & $48.69 \pm 2.05$  & $1.73 \pm 0.16$  \\
HD 202109 &Clear & $0.41\pm0.06$ &$-0.01 \pm 0.10$ & $5010 \pm 50$ & 10 & $102.1 \pm 5.2$ & $50.58 \pm 1.84$    & $2.74 \pm 0.20$   \\
HD 26886 &Clear & $0.56\pm0.10$ &$-0.30 \pm 0.10$ & $5000 \pm 100$ & 2 & $75.2 \pm 4.8$ & $51.15 \pm 4.26$& $2.06 \pm 0.26$  \\
HD 18182 &Clear & $0.40\pm0.11$ & $-0.17 \pm 0.10$ & $4900 \pm 100$ & 2 & $73.5 \pm 5.2$ & $69.52 \pm 1.28$ &  $2.93 \pm 0.30$  \\
HD 184001 &Clear & $0.59\pm0.11$ & $-0.21 \pm 0.14$ & $5000 \pm 100$ & 2 & $70.8 \pm 13.4$ & $73.72 \pm 4.46$ & $2.79 \pm 0.59$   \\
\hline
\multicolumn{9}{c}{Ba Dwarfs} \\
\hline
HD 95241 & Clear & $0.29\pm0.06$ & $-0.30 \pm 0.06$ & $5916 \pm 15$ & b & $7.1 \pm 0.3$ & $607.83 \pm 19.85$ &  $1.28 \pm 0.07$  \\
HD 124850 & Clear & $0.40\pm0.13$ &$-0.06 \pm 0.14$ & $6217 \pm 31$ & b & $9.2 \pm 0.4$ & $663.41 \pm 25.48$ & $1.52 \pm 0.10$  \\
HD 82328 & Clear & $0.26\pm0.07$ &$-0.12 \pm 0.09$ & $6371 \pm 43$ & b &$7.4 \pm 0.4$ & $791.76 \pm18.21$  & $1.34 \pm 0.08$ \\
HD 13555 & Clear & $0.30\pm0.09$ &$-0.14 \pm 0.08$ & $6485 \pm 49$ & b & $5.8 \pm 0.3$ & $1096.70 \pm 67.23$  &$1.38 \pm 0.11$  \\
HD 2454 & Clear & $0.61\pm0.03$ & $-0.34 \pm 0.04$ & $6440 \pm 100$ & d & $4.4 \pm 0.2$ & $1390.57 \pm 74.28$ &  $1.34 \pm 0.12$  \\
HD 26367 & Clear & $0.54\pm0.15$ & $-0.05 \pm 0.03$ & $6160 \pm 70$ & e & $2.8 \pm 0.2$ & $1895.62 \pm 82.56$ & $1.36 \pm 0.11$  \\
HD 87080 & Clear & $1.42\pm0.03$ & $-0.49 \pm 0.04$ & $5460 \pm 100$ & d & $7.0 \pm 1.3$ & $758.39 \pm 16.06$  & $2.09 \pm 0.42$  \\
HD 36667 & Clear & $0.36\pm0.04$ & $-0.45 \pm 0.07$ & $5776 \pm 90$ & f & $3.5 \pm 0.2$ & $1092.82 \pm 25.02$  & $1.24 \pm 0.10$ \\
HD 98991 & Marginal & $0.57\pm0.15$ & $-0.10 \pm 0.10$ & $6643 \pm 100$ & a & $15.2 \pm 0.8$ & $502.41 \pm 29.45$  & $1.51 \pm 0.14$\\
HD 48565 & Marginal &$1.37\pm0.15$ & $-0.59 \pm 0.15$ & $6030 \pm 100$ & c & $2.7 \pm 0.1$ & $1258.49 \pm 59.04$ &  $0.94 \pm 0.08$\\

HD 13551 & Marginal & $0.98\pm0.03$ & $-0.44 \pm 0.04$ & $5870 \pm 100$ & d & $2.6 \pm 0.5$ & $935.41 \pm 3.40$ &  $0.74 \pm 0.14$ \\
HD 140324 & Marginal & $0.25\pm0.15$ & $-0.36 \pm 0.07$ & $5822 \pm 90$ & f & $3.0 \pm 0.2$ & $1393.55 \pm 37.85$ & $1.33 \pm 0.10$ \\
HD 221531 & Marginal & $0.80\pm0.15$ & $-0.27 \pm 0.10$ & $6460 \pm 100$ & g & $5.4 \pm 0.4$ & $613.37 \pm 5.67$ & $0.73 \pm 0.07$ \\
\hline
\end{tabular}
\tablebib{
(1)~\cite{2013Ap.....56...57R}, (2)~\cite{2016MNRAS.459.4299D}, (3)~\cite{2007A&A...468..679S}, (4)~\cite{2003ARep...47..648A}, (5)~\cite{2013AJ....146...39K}, (6)~\cite{2024MNRAS.534.3104Y} , (7)~\cite{2018A&A...618A..32K}, (8)~\cite{2015MNRAS.446.2348K}, (9)~\cite{2019A&A...626A.127J}, 10~\cite{2018MNRAS.476..724K}, (a)~\cite{1993A&A...275..101E}, (b)~\cite{2017AJ....153...21L}, (c)~\cite{2014MNRAS.440.1095K}, (d)~\cite{2006A&A...454..895A}, (e)~\cite{2007AJ....134...96G}, (f)~\cite{2003MNRAS.340..304R}, (g)~\cite{1994A&A...281..775N}
}

\end{table*}
\subsection{Estimating $\nu_{\max}$ and $\Delta\nu$ using pySYD}
\label{validation}
\begin{figure*}[!htpb]
\sidecaption
  \includegraphics[width=10cm]{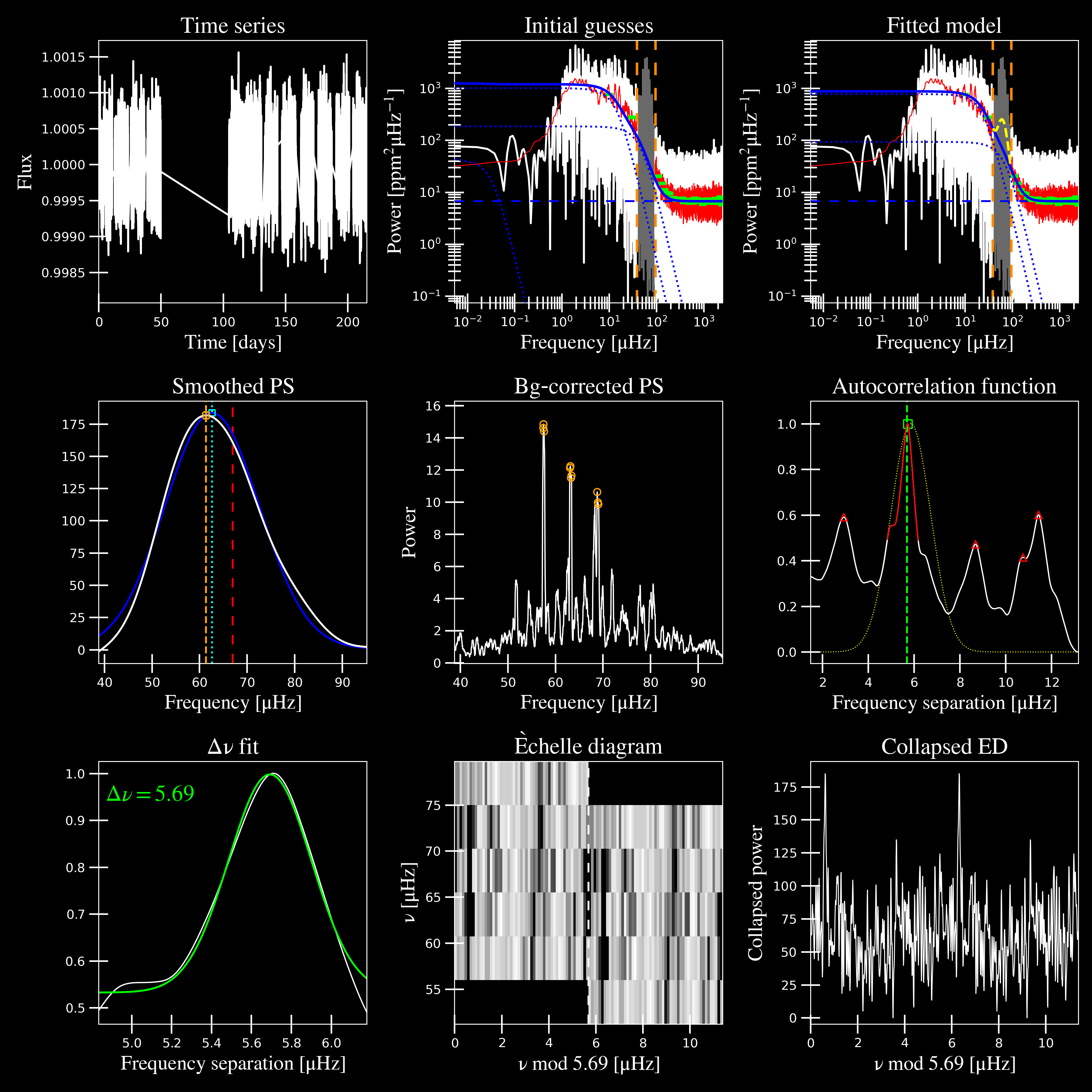}
     \caption{Estimating global asteroseismic parameters using pySYD for a Ba giant, HD 30554. \textit{Top right}: Background fit to the PSD, and a heavily smoothed version of the PSD in yellow. \textit{Centre right}: ACF and the red region show the extracted ACF peak that is used to measure $\Delta\nu$. \textit{Bottom left}:  Gaussian fit to that peak (green), with the centre of the Gaussian being the estimate of $\Delta\nu$.}
     \label{pysyd}
\end{figure*}

Figure~\ref{pysyd} shows the estimation of $\nu_{max}$ and $\Delta\nu$ using pySYD for a sample Ba giant, HD 30554. As shown in the central left panel of Figure~\ref{pysyd}, pySYD estimates $\nu_{max}$ from the peak of the smoothed, background-corrected PE (orange line). The bottom-left panel shows the extracted ACF peak in white and a fitted Gaussian in green, whose peak gives $\Delta\nu$. The estimated $\Delta\nu$ values are given in Table~\ref{delnu mass}

\begin{figure*}[!htpb]
\sidecaption
   \centering
   \includegraphics[width=8cm]{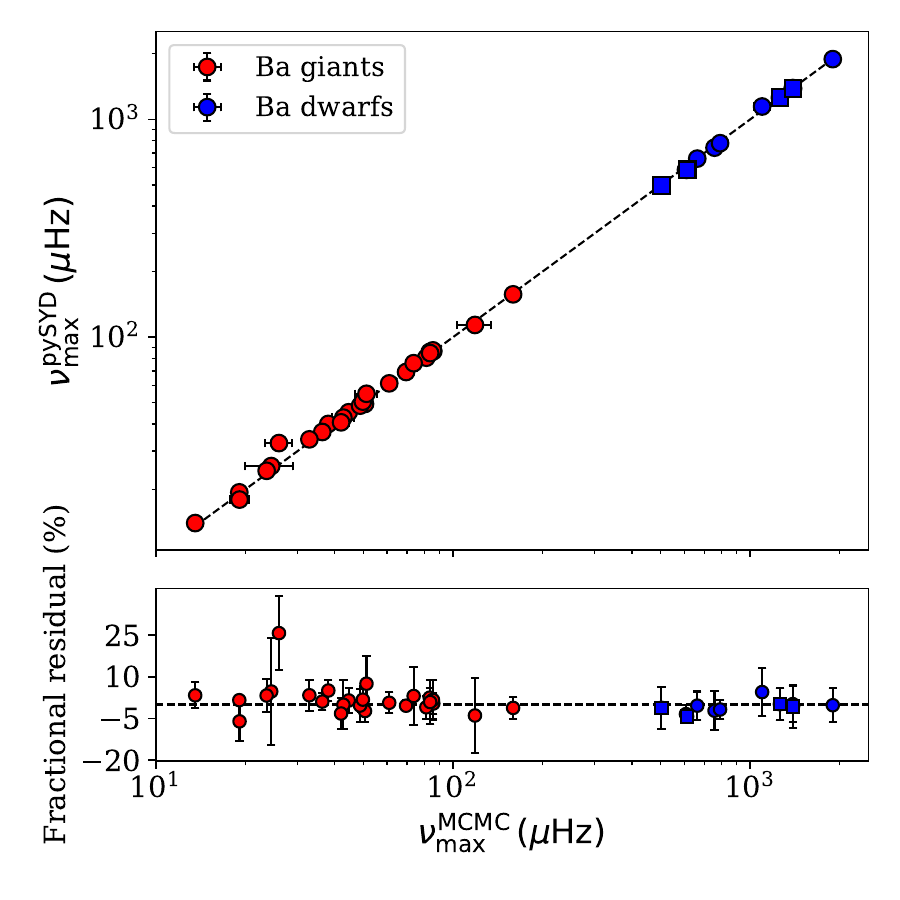}
      \caption{\textit{Top}: Comparison of $\nu_{max}$ from this work ($\nu_{max}^{MCMC}$) with pySYD values ($\nu_{max}^{pySYD}$) for the sample Ba giants (red) and dwarfs (blue). The blue squares correspond to the $\nu_{max}$ values of Ba dwarfs with marginal detections. \textit{Bottom}: Fractional residuals of $\nu_{max}$, calculated as $(\nu_{max}^{pySYD}-\nu_{max}^{MCMC})/\nu_{max}^{MCMC}$.
              }
         \label{numax compare}
   \end{figure*}

In Figure~\ref{numax compare} we compare $\nu_{max}$ estimated using the MCMC method given in Sect.~\ref{estimating numax} ($\nu_{max}^{MCMC}$) with the values obtained using pySYD ($\nu_{max}^{pySYD}$). The two approaches show overall consistency for giants and dwarfs, with the vast majority of targets having fractional residuals,  $(\nu_{max}^{pySYD}-\nu_{max}^{MCMC})/\nu_{max}^{MCMC}$ below $10\%$.
The mean fractional residual is $3.2\%$ with a $1\sigma$ deviation of $4.8\%$, indicating strong agreement between the two methods~\citep{2024ApJS..271...17Z}.

\subsection{\texorpdfstring{Estimating $\Delta P$}
{Estimating Delta_P}}
\begin{figure*}
\sidecaption
    \centering
    \includegraphics[width=9cm]{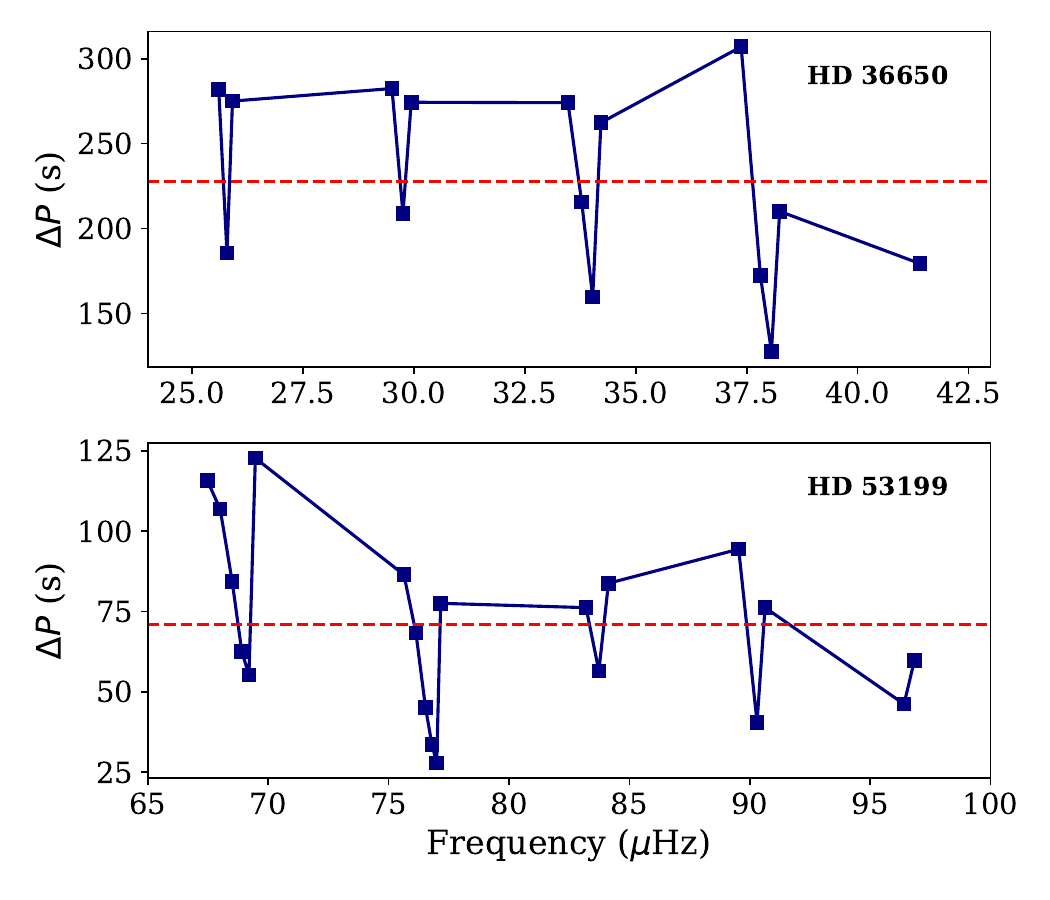}
    \caption{Representative $\Delta P$ measurements for Ba giants. The dashed red lines represent the average of the period spacing between consecutive $l=1$ modes.}
    \label{delp}
\end{figure*}

Figure~\ref{delp} shows the $\Delta P$ measurement for representative Ba giants. As explained in Sect.~\ref{evs}, the period spacing between consecutive $l=1$ modes is shown in blue, and their corresponding average value ($\Delta P$) is shown by the dashed red line.

\subsection{Asteroseismic masses}
With measured $\Delta\nu$ for 14 Ba giants and 5 Ba dwarfs, we estimated the seismic masses, $M_1$, $M_2$, $M_3$ and $M_4$ from the scaling relations given by Eqs.~\ref{eq8},~\ref{eq9},~\ref{eq10}, and~\ref{eq11} respectively in Sect.~\ref{scaling rel}. We present the seismic masses together with the average masses of this sub-sample in Table~\ref{delnu mass}.
\begin{table*}[!htpb]
\centering
\caption{$\Delta\nu$, $\Delta P$, and asteroseismic masses for the sample of Ba giants and dwarfs.}
\label{delnu mass}
\begin{tabular}{lccccccccccccccc}
\hline\hline
Object  & $f_{\Delta\nu}$ & $\Delta\nu~(\mu$Hz) & $\Delta P$~(s) & $M_1~(\rm{M}_\odot)$ & $M_2~(\rm{M}_\odot)$ & $M_3~(\rm{M}_\odot)$ & $M_4~(\rm{M}_\odot)$ & $M_{av}~(\rm{M}_\odot)$ \\
\hline
\multicolumn{9}{c}{Ba Giants} \\
\hline
HD 36650 & $1.00 \pm 0.02$ & $4.07 \pm 0.17$ & $228 \pm 14$ & $1.12 \pm 0.23$ & $1.01 \pm 0.33$ & $1.04 \pm 0.21$ & $1.09 \pm 0.17$ & $1.07 \pm 0.02$ \\
HD 22772 &  $0.98 \pm 0.02$ & $4.82 \pm 0.18$ & $103 \pm 9$ & $1.80 \pm 0.38$ & $1.66 \pm 0.30$ & $1.70 \pm 0.18$ & $1.77 \pm 0.27$ & $1.73 \pm 0.03$ \\
HD 24035  & $0.97 \pm 0.02$ & $5.17 \pm 0.04$ & $76 \pm 5$ & $1.24 \pm 0.15$ & $0.67 \pm 0.22$ & $0.82 \pm 0.17$ & $1.09 \pm 0.11$ & $0.96 \pm 0.11$ \\
HD 30554 &  $0.98 \pm 0.02$ & $5.69 \pm 0.07$ & $86 \pm 6$ & $1.71 \pm 0.21$ & $1.65 \pm 0.27$ & $1.67 \pm 0.16$ & $1.70 \pm 0.15$ & $1.68 \pm 0.01$ \\
HD 65314 &  $0.99 \pm 0.02$ & $6.92 \pm 0.05$ & $36 \pm 3$ & $2.08 \pm 0.32$ & $2.63 \pm 0.84$ & $2.43 \pm 0.51$ & $2.18 \pm 0.28$ & $2.33 \pm 0.11$ \\
HD 53199 &  $0.99 \pm 0.02$ & $7.11 \pm 0.08$ & $71 \pm 6$ & $1.99 \pm 0.26$ & $2.57 \pm 0.42$ & $2.36 \pm 0.24$ & $2.10 \pm 0.21$ & $2.25 \pm 0.11$ \\
HD 107541 & $0.98 \pm 0.02$ & $13.54 \pm 1.15$ & $38 \pm 3$ & $1.02 \pm 0.38$ & $1.05 \pm 0.37$ & $1.04 \pm 0.21$ & $1.02 \pm 0.27$ & $1.03 \pm 0.01$ \\
HD 32712 & $0.98 \pm 0.02$ & $3.06 \pm 0.06$ & -- & $1.12 \pm 0.20$ & $1.18 \pm 0.37$ & $1.16 \pm 0.24$ & $1.13 \pm 0.16$ & $1.15 \pm 0.01$ \\
HD 83548 &  $1.01 \pm 0.02$ & $3.92 \pm 0.18$ & -- & $3.13 \pm 1.03$ & $3.77 \pm 1.21$ & $3.55 \pm 0.77$ & $3.25 \pm 0.84$ & $3.43 \pm 0.12$ \\
HD 116713 &  $0.99 \pm 0.02$ & $4.28 \pm 0.06$ & -- & $1.35 \pm 0.21$ & $0.84 \pm 0.25$ & $0.99 \pm 0.19$ & $1.23 \pm 0.16$ & $1.10 \pm 0.10$ \\
HD 199394 &$1.00 \pm 0.02$ & $6.79 \pm 0.13$ & -- & $2.62 \pm 0.48$ & $1.93 \pm 0.30$ & $2.14 \pm 0.21$ & $2.46 \pm 0.35$ & $2.29 \pm 0.14$ \\
HD 39778 & $1.00 \pm 0.02$ & $7.24 \pm 0.23$ & -- & $2.09 \pm 0.39$ & $3.50 \pm 1.10$ & $2.94 \pm 0.60$ & $2.32 \pm 0.34$ & $2.71 \pm 0.27$ \\
HD 160507 & $1.00 \pm 0.02$ & $7.43 \pm 0.06$ & -- & $1.92 \pm 0.30$ & $1.94 \pm 0.20$ & $1.93 \pm 0.14$ & $1.92 \pm 0.23$ & $1.93 \pm 0.01$ \\
HD 29685 & $1.00 \pm 0.02$ & $9.63 \pm 0.46$ & -- & $1.69 \pm 0.76$ & $1.51 \pm 0.35$ & $1.57 \pm 0.29$ & $1.65 \pm 0.58$ & $1.61 \pm 0.03$ \\
\hline
\multicolumn{9}{c}{Ba Dwarfs} \\
\hline
HD95241 & $1.01 \pm 0.02$  & $36.45 \pm 0.80$ &  & $1.55 \pm 0.24$ & $1.16 \pm 0.11$ & $1.28 \pm 0.07$ & $1.47 \pm 0.17$ & $1.36 \pm 0.08$ \\
HD124850  &$0.99 \pm 0.02$& $34.65 \pm 2.63$  & &$2.43 \pm 0.82$ & $1.20 \pm 0.21$ & $1.52 \pm 0.10$ & $2.11 \pm 0.51$ & $1.81 \pm 0.24$ \\
HD82328  &$0.98 \pm 0.02$ & $44.68 \pm 0.57$ & & $1.51 \pm 0.18$ & $1.27 \pm 0.12$ & $1.34 \pm 0.08$ & $1.46 \pm 0.13$ & $1.40 \pm 0.05$ \\
HD13555 &$0.97 \pm 0.02$& $56.82 \pm 2.14$ && $1.49 \pm 0.38$ & $1.32 \pm 0.16$ & $1.38 \pm 0.11$ & $1.45 \pm 0.28$ & $1.41 \pm 0.03$ \\
HD2454 & $0.96 \pm 0.02$ & $68.22 \pm 3.41$ & & $1.42 \pm 0.38$ & $1.31 \pm 0.21$ & $1.34 \pm 0.12$ & $1.39 \pm 0.28$ & $1.36 \pm 0.02$ \\
\hline
\end{tabular}
\tablefoot{
$M_{av}$~ is the average mass from $M_1,~M_2,~M_3,~M_4$. This table contains a subset of 14 giants and 5 dwarfs for which $\Delta\nu$ estimation was possible. The full sample is given in Table~\ref{gba dba param}
}
\end{table*}
\onecolumn
\section{Behaviour of $s$-process abundance with [Fe/H]}
\begin{figure*}[h]
    \centering
    \includegraphics[width=12cm]{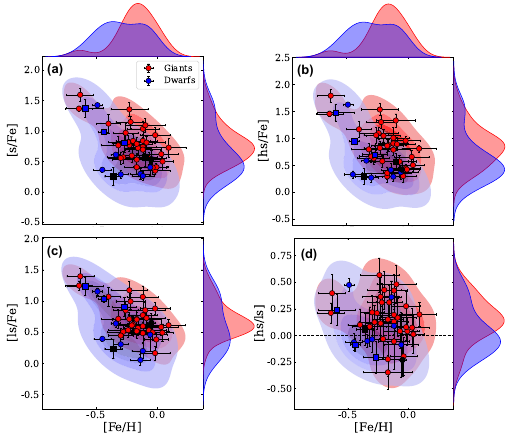}
    \caption{Behaviour of $s$-process abundance with respect to [Fe/H]. \textit{Panels (a), (b), (c), and (d)}: [s/Fe], [hs/Fe], [ls/Fe], and [hs/ls] versus [Fe/H]. Symbols are the same as Figure~\ref{sfe_hsfe_obs}.}
    \label{s_feh}
\end{figure*}

In Figure~\ref{s_feh} we show the $s$-process abundance versus metallicity of our Ba star sample. As explained in Sect.~\ref{s observed}, we observe the expected anti-correlation of $s$-process abundance with [Fe/H]. However, we note that the bulk of the sample has a narrow range of metallicity ($\approx0.15\pm0.2$~dex).

\FloatBarrier
\section{Modelling Ba stars}
In this section, we describe some of the details of stellar models of Ba stars together with the role of accretion on the extent of TH mixing.

\subsection{\texorpdfstring{Modelling $s$-process abundances in Ba stars using FRUITY yields}
{Modelling s-process abundances in Ba stars using FRUITY yields}}
\label{appendix fruity mesa}
\begin{figure*}
\sidecaption
  \includegraphics[width=12cm]{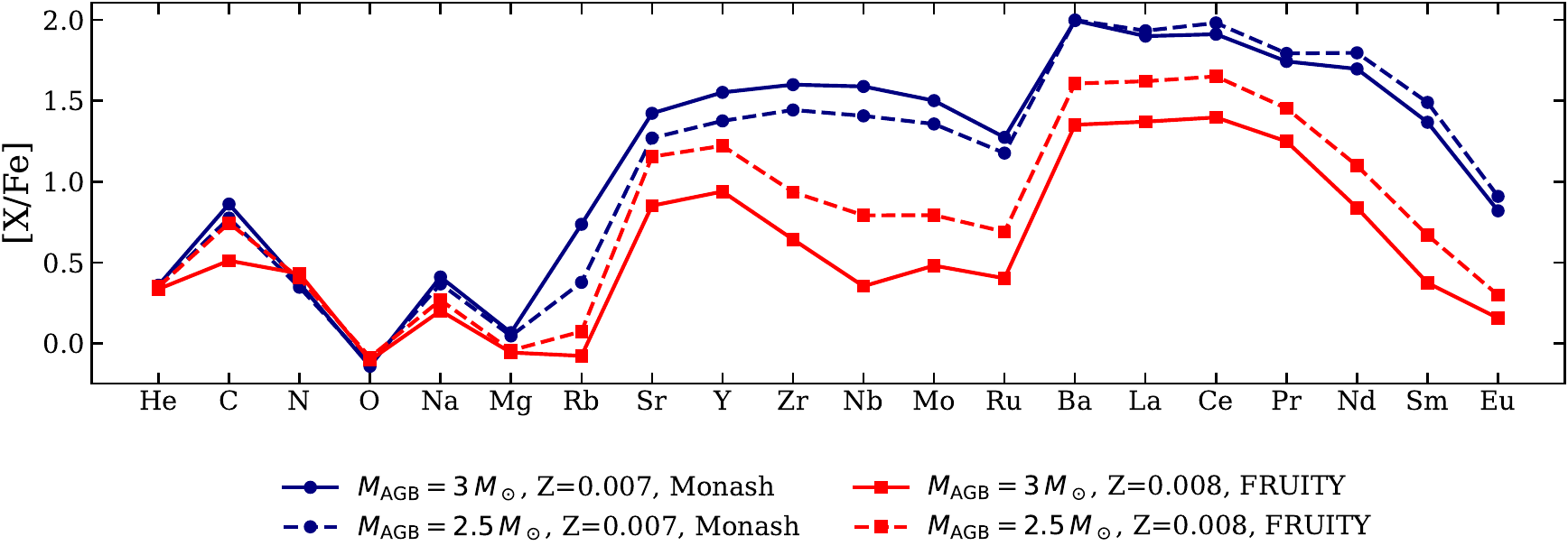}
     \caption{Comparison of stellar yields ([X/Fe]) for different elements from FRUITY (red) and Monash (blue) AGB models. The solid lines correspond to $3~\rm{M}_\odot$, and dashed lines correspond to $2.5~\rm{M}_\odot$ AGB models. FRUITY yields are considerably lower than Monash yields, especially for heavy elements.}
     \label{fruity_Monash}
\end{figure*}

In Table~\ref{FRUITY model} we present the masses of the modelled Ba stars and the corresponding [s/Fe] and metallicity of the FRUITY AGB yields used. We did not consider any FRUITY models above $4~\rm{M}_\odot$~as the $s$-process production is insufficient to produce a Ba star after pollution. As mentioned in Sect.~\ref{model description}, FRUITY yields are considerably lower than Monash yields, especially for heavy elements. This is evident in Figure~\ref{fruity_Monash}, which shows the comparison between the AGB yields of light and heavy elements from FRUITY and Monash AGB yields for $2.5~M_\odot$ and $3~M_\odot$ models with similar metallicities.

\begin{figure*}
    \centering
    \includegraphics[width=\hsize]{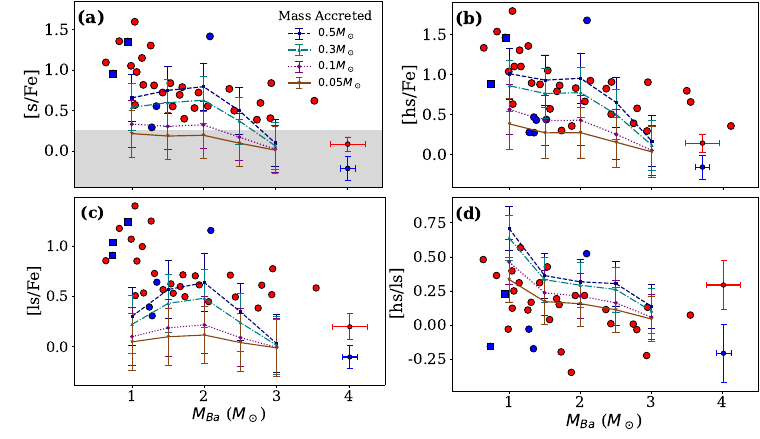}
    \caption{Comparing $s$-process abundance ratios from MESA models using accreted AGB composition from FRUITY AGB yields with observed values for Ba giants (red filled circles), Ba dwarfs with clear detections (blue filled circles), and marginal detections (blue filled squares). The representative error bar (median error) is given in red for Ba giants and blue for Ba dwarfs, respectively, in the bottom right of each panel. Note that the AGB donors are assumed to be $\sim1.3\times M_{\rm Ba}$ as given in Table~\ref{FRUITY model} and that all models include TH mixing. \textit{Panels (a), (b), (c), and (d)}: [s/Fe], [hs/Fe], [ls/Fe], and [hs/ls] versus final mass of the Ba star after accretion ($M_{\rm Ba}$). The grey shaded region in panel (a) denotes [s/Fe] $\leq0.25~$dex. The model lines are the same as Figure~\ref{mesa}.
}
     \label{mesa_fruity}
\end{figure*}

Following the same methodology of Sect.~\ref{model description}, we present the comparison of modelled $s$-process abundances using FRUITY yields with the observed abundances in Figure~\ref{mesa_fruity} (see Sect.~\ref{sfe fruity} for details). Since only one model is available for each mass for $Z=0.008$, the model error bars correspond to the average of the standard deviations of abundances across all relevant mass models and metallicity.

\begin{table}
\caption{Mass ($M_{\rm AGB})$, metallicity (Z) and [s/Fe] of FRUITY AGB models and the final mass of Ba star $(M_{\rm Ba})$ used in the MESA models.}          
\label{table:A2}     
\centering                       
\begin{tabular}{c c c c}     
\hline\hline               
$M_{\rm Ba} (\rm{M}_\odot)$ & $M_{\rm AGB} (\rm{M}_\odot)$ & Z& [s/Fe]   \\    
\hline                       
1 & 1.5 & 0.003 & $0.89 \pm 0.30$\\
1.5 & 2 & 0.008 & $1.11\pm0.30$\\
2 & 2.5 & 0.008 & $1.28 \pm0.30$\\
2.5 & 3 & 0.008 & $1.01 \pm 0.30$\\
3 & 4 & 0.008 & $0.33 \pm 0.30$\\
\hline
\label{FRUITY model}
\end{tabular}
\end{table}

\subsection{Light elements in Ba stars and comparison with stellar models incorporating FRUITY yields}
\label{appendixB}
\begin{figure*}
    \centering
    \includegraphics[width=\hsize]{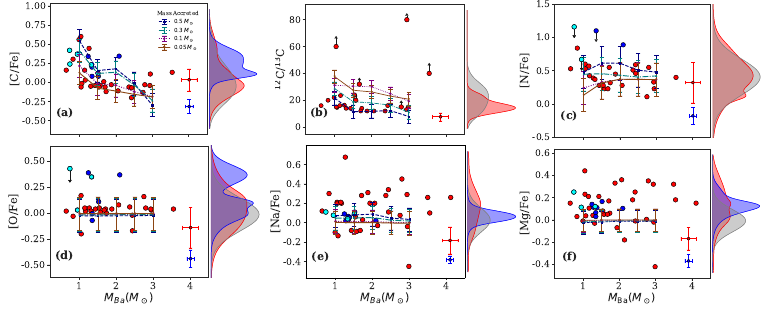}
    \caption{Comparing light elements from MESA models with accreted AGB composition from FRUITY AGB yields with observed values for Ba giants (red) and dwarfs (blue). The Ba dwarfs with marginal detections are shown in cyan. The up arrows correspond to stars for which the lower limit of that abundance is available. Down arrows correspond to the upper limit of that abundance. The representative error bar for Ba giants and Ba dwarfs is shown in red and blue, respectively, in the bottom-right of each panel. Note that the AGB donors are $\sim1.3\times M_{\rm Ba}$ as given in Table~\ref{FRUITY model}, and all the models include TH mixing. The model lines and KDEs are the same as Figure~\ref{mesa_light}.}
    \label{mesa_light_fruity}
\end{figure*}

Table~\ref{gba light} shows the average light element abundances of the sample Ba giants and dwarfs collected from the literature. Contrary to the $s$-process abundances, our stellar models incorporating AGB composition from FRUITY AGB yields can satisfactorily reproduce the observed abundances of C, N and $^{12}$C/$^{13}$C ratio, except for a few outliers, as shown in Figure~\ref{mesa_light_fruity}. We note that models incorporating FRUITY yields produce slightly higher [C/Fe] by $\sim0.1$~dex relative to those of Monash yields. However, that is consistent with typical abundance uncertainties. In this case, we also observe a similar overabundance of observed [O/Fe], [Na/Fe], and [Mg/Fe] with respect to model predictions as explained in Sect.~\ref{Light_obs_model}. 
\label{Fruity light}

\begin{sidewaystable*}
\centering
\scriptsize
\caption{Average abundances of light elements (C to Mg)  in Ba giants (\textit{top}) and Ba dwarfs (\textit{bottom}) collected from the literature along with their references in the Ref column.}
\begin{tabular}{lcccccccccccccc} 
\hline\hline
Object & $M_3 (\rm{M}_\odot)$ & $\langle$ [C/Fe] $\rangle$ & Ref & $\langle ^{12}$C/$^{13}$C$\rangle$ & Ref & $\langle$ [N/Fe] $\rangle$ & Ref & $\langle$ [O/Fe]$\rangle$ & Ref & $\langle$ [Na/Fe]$\rangle$ & Ref & $\langle$ [Mg/Fe]$\rangle$ & Ref \\
\hline
\multicolumn{14}{c}{Ba Giants} \\
\hline
HD 66812 & 4.11$\pm$1.12 & -- & -- & -- & -- & -- & -- & -- & -- & 0.26 $\pm$ 0.11 & 1 & 0.15 $\pm$0.03 & 1  \\
HD 32712 & 1.16 $\pm$ 0.24 & 0.03 $\pm$ 0.13 & 2,3 & 24 $\pm$ 3 & 2,3 & 0.42$\pm$0.19 & 2,3 & 0.02$\pm$0.18 & 2,3 & 0.20       $\pm$ 0.12 & 2,4 & 0.11 $\pm$ 0.12 & 2,4\\
HD 83548 & 3.55 $\pm$ 0.77 & 0.13 $\pm$ 0.07 & 3,5,6 & $\geq 40\pm6$ & 3 & 0.40 $\pm$ 0.18 & 3,6 & 0.04 $\pm$0.12 & 3,6 & 0.10 $\pm$0.14 & 4 & 0.18 $\pm$0.10 & 4 \\
HD 36650 &  1.04 $\pm$ 0.21 & $-0.06\pm0.12$ & 2,3 & 15$\pm$ 3 & 3 & 0.53$\pm$0.18& 2,3 & $-0.20\pm0.17$ & 2,3 & 0.20$\pm$0.06 & 2,4 & 0.23 $\pm$   0.12 & 2,4 \\
HD 116713 & 0.99 $\pm$  0.19 & 0.56 $\pm$ 0.10 & 5 & 22 $\pm$   3 & 7,8 & -- & -- & -- & -- & $-0.10\pm0.10$ & 9 & $-0.01 \pm 0.16$ & 9 \\
HD 22772 & 1.70 $\pm$   0.18 & $-0.06 \pm       0.16$ & 3 & 12 $\pm$ 3 & 3 & 0.55 $\pm$  0.30 & 3 & 0.04 $\pm$   0.21 & 3 & 0.28 $\pm$   0.14 & 4 & 0.17 $\pm$    0.10 & 4\\
HD 90167 & 1.73 $\pm$   0.16 & -0.12$\pm$0.14 & 3 & 14$\pm$6 & 3 & 0.32$\pm$0.32 & 3 & 0.00 $\pm$        0.20 & 3 & 0.03 $\pm$0.14 & 4 & 0.06 $\pm$0.10 & 4 \\
HD 24035 & 0.82$\pm$0.17 & 0.30$\pm$    0.09 &2,3,10 & 20$\pm$5 & 10 & 0.84 $\pm$0.13 & 2,3,10 & $-0.03 \pm 0.15$ & 3,10 & 0.30$\pm$        0.13 & 2,4 & 0.03 $\pm$0.08 & 2,4,10\\
HD 30554 & 1.67 $\pm$0.16 & 0.01 $\pm$0.17 & 3 & $\geq32\pm3$ & 3 & 0.23$\pm$0.31 & 3 & 0.03 $\pm$0.21 & 3 & $-0.08 \pm   0.14$ & 4 & 0.05$\pm$0.10 & 4 \\
HD 199394 & 2.14 $\pm$  0.21 & -- & --& -- & -- & -- & -- &-- & -- & $-0.03\pm0.07$ & 11,12 & $-0.18\pm     0.10$ & 12 \\
HD 65314 & 2.43 $\pm$0.51 & $-0.07      \pm0.07$ & 13 & -- & -- & 0.46$\pm$0.18 & 13 & $-0.17 \pm       0.09$ & 13 & 0.11 $\pm$ 0.09 & 13 & $-0.01 \pm  0.08$ & 13\\
HD 53199 & 2.36 $\pm$   0.24 & $-0.06 \pm       0.16$ & 3 & 12 $\pm$    3 & 3 & 0.51 $\pm$        0.30 & 3 &0.06 $\pm$    0.21 & 3 &0.28 $\pm$    0.14 & 4 & 0.18 $\pm$        0.10 & 4\\
HD 39778 & 2.94 $\pm$   0.60 & 0.11 $\pm$       0.16 & 3 & $\geq80\pm   3$ & 3 & 0.23 $\pm$        0.30 & 3 & 0.03 $\pm$   0.20 & 3 & $-0.04 \pm   0.14$ & 4 & $-0.02 \pm        0.10$ & 4\\
HD 160507 & 1.93 $\pm$  0.14 & -- & -- & -- & -- & -- & -- & -- & -- & 0.20 $\pm$   0.10 & 14 & $-0.07\pm   0.1$ & 14\\
HD 29685 & 1.57 $\pm$0.29 & $-0.04\pm   0.17$ & 3 & $\geq20\pm  3$& 3 & 0.23 $\pm$   0.31 & 3 & 0.01 $\pm$   0.21 & 3 & $-0.09 \pm   0.14$ & 4 & 0.05 $\pm$   0.10 & 4\\
HD 107541 & 1.04 $\pm$  0.21 & 0.60 $\pm$       0.14 & 3 & $\geq60\pm   6$ & 3 & 0.6 $\pm$ 0.32 & 3 & 0.17 $\pm$   0.20 & 3 & 0.21 $\pm$   0.14 & 4 & 0.44 $\pm$    0.10 & 4\\
HD 16458 & 1.26 $\pm$   0.09 & 0.44 $\pm$       0.08 & 5,15 & 16 $\pm$  3 &  15,16,17 & 0.39 $\pm$        0.20 & 15 &  0.05 $\pm$ 0.18 & 15 & 0.68 $\pm$   0.09 & 15,18 & 0.04 $\pm$       0.10 & 18\\
HD 111315 & 2.43 $\pm$  0.49 & $-0.20\pm        0.13$ & 3 & 16$\pm$3 & 3 & 0.59 $\pm$    0.29 & 3 & $-0.02 \pm   0.18$ & 3 & 0.34 $\pm$  0.14 & 4 & 0.25 $\pm$    0.10 & 4 \\
BD-083194 & 0.64 $\pm$  0.07 & 0.16 $\pm$       0.19 & 3 & 16$\pm$3 & 3 & 0.53 $\pm$      0.32 & 3 & 0.02 $\pm$   0.23 & 3 & 0.12 $\pm$   0.14 & 4 & 0.15 $\pm$    0.10 & 4 \\
CD-611941 & 1.54 $\pm$  0.44 & $-0.06 \pm       0.17$ &3 & 24$\pm$      3& 3 & 0.28        $\pm$0.31 & 3 & 0.05 $\pm$      0.21 & 3 & 0.06 $\pm$   0.14 & 4 &0.08 $\pm$ 0.10 & 4\\
HD 104979 & 1.07 $\pm$  0.10 & $-0.01 \pm       0.06$ & 5,19,20 & 14 $\pm$      3 & 8,19,21,22 & -- & -- & -- & -- & -0.14 $\pm$0.07 & 9,19 & 0.04 $\pm$0.06 & 9,19,20\\
HD 213084 & 1.31 $\pm$  0.28 & $-0.05 \pm0.18$ & 3 & 14 $\pm$   6 & 3 & 0.37 $\pm$0.34 & 3 & 0.02 $\pm$      0.22 & 3 & 0.04 $\pm$   0.14 & 4 & $-0.03 \pm     0.10$ & 4\\
HD 50075 & 1.43 $\pm$   0.34 & $-0.04 \pm       0.15$ & 3 & 14$\pm$3 & 3 & 0.28 $\pm$    0.30 & 3 & 0.04$\pm$0.20 & 3 & $-0.08 \pm       0.14$ & 4 & 0.21 $\pm$    0.10 & 4 \\
HD 177192 & 1.87 $\pm$  0.39 & $-0.03 \pm0.22$ & 3 & 12 $\pm$   3 & 3 & 0.45 $\pm$0.34 & 3 & 0.04 $\pm$      0.26 & 3 & 0.31 $\pm$   0.14 & 4 & 0.33 $\pm$   0.10 & 4\\
HD 20394 & 1.14 $\pm$   0.12 & 0.16 $\pm$ 0.15 & 3 & 16 $\pm$   3 & 3 & 0.57 $\pm$   0.33 & 3 & 0.05 $\pm$   0.20 & 3 & 0.21 $\pm$   0.10 & 4 & 0.21$\pm$0.08 & 4 \\
HD 101013 & 2.98 $\pm$  0.27 & -- & -- & 15 $\pm$       4 & 7,17 & -- & --& -- & -- & $-0.45 \pm    0.10$ & 12 & $-0.42 \pm 0.10$ & 12 \\
HD 62017 & 3.49 $\pm$   0.55 & -- & --& -- & --& -- & -- & -- & -- & 0.26 $\pm$0.14 & 4 & 0.32 $\pm$      0.10 & 4 \\
HD 202109 & 2.74 $\pm$  0.20 & $-0.13\pm0.08$ & 23,24 & -- & -- & 0.11$\pm$0.05 & 24 & 0.16 $\pm$       0.06 & 24 & 0.08 $\pm$  0.07 & 9,23 & 0.005 $\pm$       0.045 & 9,23\\
HD 26886 & 2.06 $\pm$   0.26 & $-0.11\pm0.14$ & 3 & 12 $\pm$    3 & 3 & 0.28 $\pm$   0.29 & 3 & 0.08 $\pm$   0.19 & 3 & 0.20 $\pm$        0.14 & 4 & 0.36 $\pm$    0.10 & 4\\
HD 18182 & 2.93 $\pm$   0.30 & $-0.04\pm0.14$ & 3 & 13 $\pm$    3 & 3 & 0.33 $\pm$   0.29 & 3 & 0.04 $\pm$   0.19 & 3 & 0.29 $\pm$   0.14 & 4 & 0.32$\pm$0.10 & 4
\\
HD 184001 & 2.79 $\pm$  0.59 & $-0.03\pm0.17$ & 3 & $\geq 15\pm3$ & 3 & 0.55 $\pm$   0.31 & 3 & 0.05 $\pm$   0.21 & 3 & 0.45 $\pm$   0.14 & 4 & 0.35$\pm$0.10 & 4\\
\hline
\multicolumn{14}{c}{Ba Dwarfs} \\
\hline
HD 95241 & 1.28 $\pm$   0.07 & -- & -- & -- & -- & -- & -- & -- & -- & 0.02$\pm$0.03 & 26,27,30 & 0.11 $\pm$    0.06 & 26,27,30\\
HD 124850 & 1.52 $\pm$  0.10 & 0.08 $\pm$       0.04 & 28 & -- & --& -- & -- & 0.11 $\pm$ 0.04 & 28 & 0.11 $\pm$  0.08 &26,29,30 & 0.16 $\pm$     0.12 & 26,30 \\
HD 82328 & 1.34 $\pm$   0.08 & 0.08 $\pm$       0.09& 31 &-- & -- & -- & -- & $0.00 \pm0.06$ & 31,32,33 & 0.03 $\pm$  0.06 & 26,27,30 & 0.03$\pm$0.11 & 26,27,30\\
HD 13555 & 1.38 $\pm$   0.11 & -- & -- & -- & -- &-- & --& -- & -- & 0.05 $\pm$   0.05 & 26,30,34 & 0.11 $\pm$    0.10 & 26,30,34\\
HD 2454 & 1.34 $\pm$    0.12 & 0.20 $\pm$       0.07 & 35,36 & -- & --& $<1.1 \pm     0.13$ & 35 & $-0.07\pm  0.11$ & 35 & 0.06 $\pm$ 0.04& 34,35 & 0.06 $\pm$   0.08 & 34,35 \\
HD 26367 & 1.36 $\pm$   0.11 & -- & --& -- & --& -- & -- & -- & -- & -- & -- & -- & --\\
HD 87080 & 2.09 $\pm$   0.42 & 0.34 $\pm$       0.08 & 35,37,38 & -- & -- & 0.89 $\pm$    0.10 & 37 & 0.37 $\pm$  0.10 & 37 & 0.02 $\pm$  0.08 & 35,37,38 & 0.10 $\pm$    0.05 & 35,37,38 \\
HD 36667 & 1.24 $\pm$   0.10 & 0.34 $\pm$       0.14 & 39 & -- & -- & -- & -- & 0.39 $\pm$       0.08 & 27,39 & 0.09 $\pm$       0.03 & 27,39 & 0.14 $\pm$   0.03 & 27,39\\
HD 98991 & 1.51 $\pm$   0.14 & 0.12     $\pm$0.04 & 28 & -- & -- & -- & -- & -- & -- & 0.09 $\pm$  0.06 & 27 & -- & -- \\
HD 48565 & 0.94 $\pm$   0.08 & 0.37 $\pm$       0.09 & 35 & -- & -- & $<        0.67\pm 0.13 $ & 35 & 0.03$\pm$0.11 & 35 & 0.08 $\pm$        0.04 & 18,26,35 & 0.11 $\pm$    0.05 & 18,26,35\\

HD 13551 & 0.74 $\pm$   0.14 & 0.24 $\pm$       0.09 & 35 & -- & -- & $<        1.16\pm 0.13$ & 35 & $<       0.43\pm 0.11$ & 35 & 0.11 $\pm$ 0.04 & 35 & 0.25 $\pm$  0.06 & 35\\
HD 140324 & 1.33 $\pm$  0.10 & 0.23 $\pm$       0.14 & 39 & -- & -- & -- & -- & 0.35 $\pm$       0.08 & 27,39 & 0.04 $\pm$       0.03 & 27,39 & 0.12 $\pm$   0.04 & 27,39,40\\
HD 221531 & 0.73 $\pm$  0.07 & 0.42 $\pm$       0.10 & 41 & -- & -- & -- & -- & -- & --& -- & -- & -- & --\\

\hline
\end{tabular}
\label{gba light}
\tablebib{
(1)~\citealt{2013Ap.....56...57R}, (2)~\citealt{2020MNRAS.492.3708S}, (3)~\citealt{2025AJ....170..259R}, (4)~\citealt{2016MNRAS.459.4299D}, 
(5)~\citealt{2007ApJ...658..367K}, (6)~\citealt{1992A&A...262..216B}, (7)~\citealt{1979ApJ...227..209T}, (8)~\citealt{1981ApJ...247.1052S}, (9)~\citealt{2007A&A...468..679S}, (10)~\citealt{2010A&A...509A..93M}, (11)~\citealt{2002A&A...389..519A}, (12)~\citealt{1994A&A...283..937Z}, (13)~\citealt{2013AJ....146...39K}, (14)~\citealt{2024MNRAS.534.3104Y}, (15)~\citealt{2018A&A...618A..32K}, (16)~\citealt{1984A&A...132..326S}, (17)~\citealt{1988IAUS..132..593J}, (18)~\citealt{2014MNRAS.440.1095K}, (19)~\citealt{2015MNRAS.446.2348K}, (20)~\citealt{2024A&A...691A.128D}, (21)~\citealt{1976ApJ...210..694T}, (22)~\citealt{1978ApJ...223L..31D}, (23)~\citealt{2018MNRAS.476..724K}, (24)~\citealt{2016A&A...586A.151M}, (25)~\citealt{2003A&A...397..257L}, (26)~\citealt{2017AJ....153...21L}, (27)~\citealt{2005A&A...438..139S}, (28)~\citealt{1999A&A...342..426G}, (29)~\citealt{2004A&A...423..683S}, (30)~\citealt{1993A&A...275..101E}, (31)~\citealt{2006AJ....131.3069L}, (32)~\citealt{1999AJ....117..492B}, (33)~\citealt{2013ApJ...764...78R}, (34)~\citealt{2014A&A...562A..71B}, (35)~\citealt{2006A&A...454..895A}, (36)~\citealt{2006MNRAS.367.1181B}, (37)~\citealt{2019MNRAS.486.3266P}, (38)~\citealt{2003A&A...402.1061P}, (39)~\citealt{2003MNRAS.340..304R}, (40)~\citealt{2004AJ....128.1177V}, (41)~\citealt{1994A&A...281..775N}
}

\end{sidewaystable*}

\subsection{Accretion and TH mixing}
\label{Kepp_acc_appendix}
\begin{figure}[!htpb]
    \centering
    \begin{subfigure}{0.5\columnwidth}
        \includegraphics[width=\hsize]{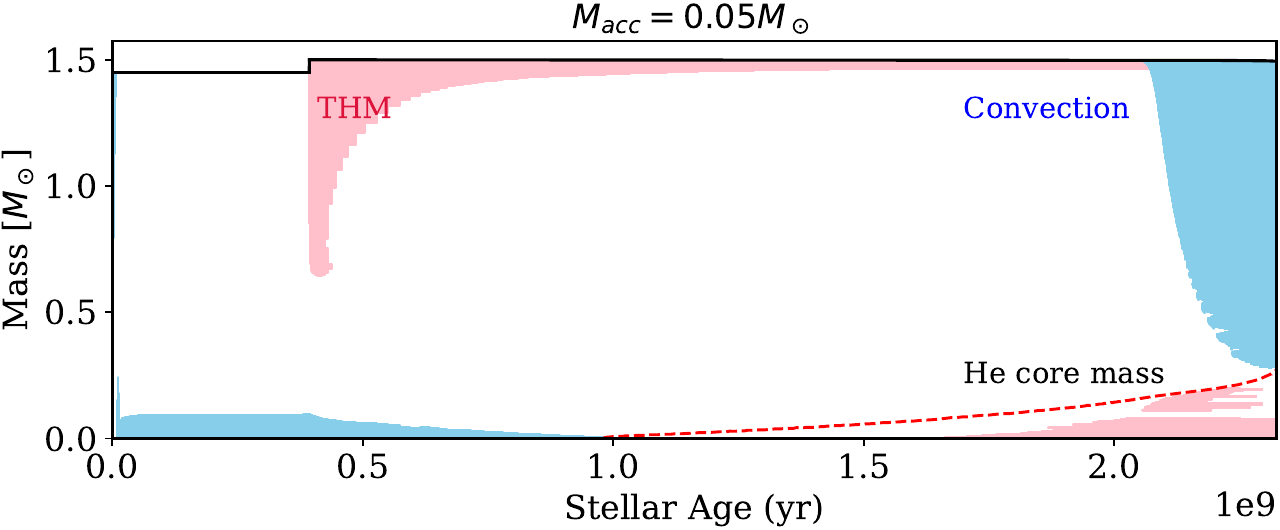}
        \caption{}
        \label{Kipp0.05}
    \end{subfigure}\\
    
    \begin{subfigure}{0.5\columnwidth}
        \includegraphics[width=\hsize]{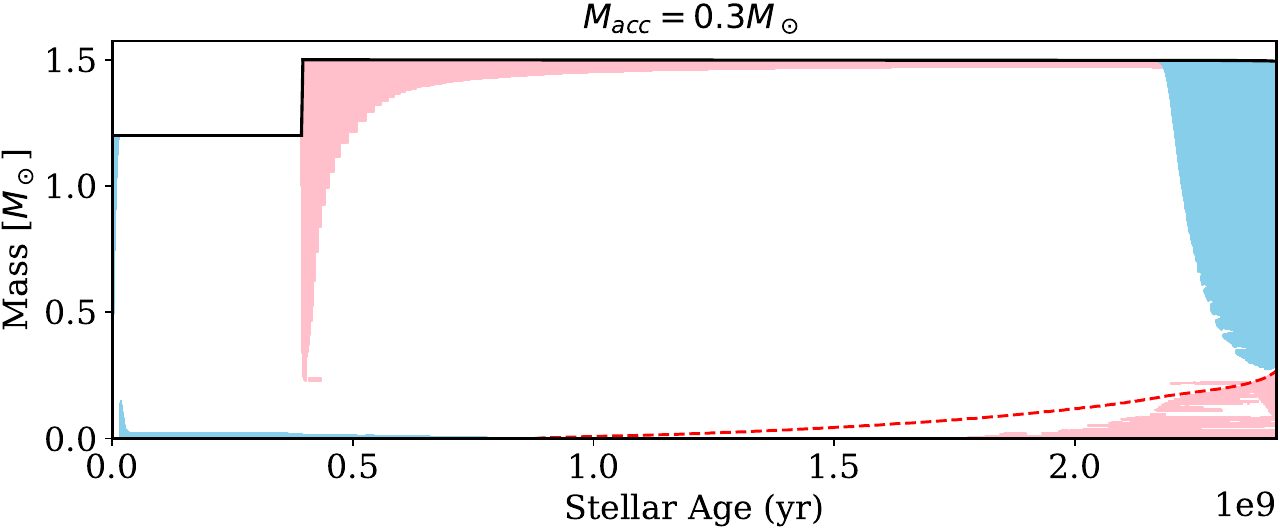}
        \caption{}
        \label{Kipp0.3}
    \end{subfigure}\\

    \begin{subfigure}{0.5\columnwidth}
        \includegraphics[width=\hsize]{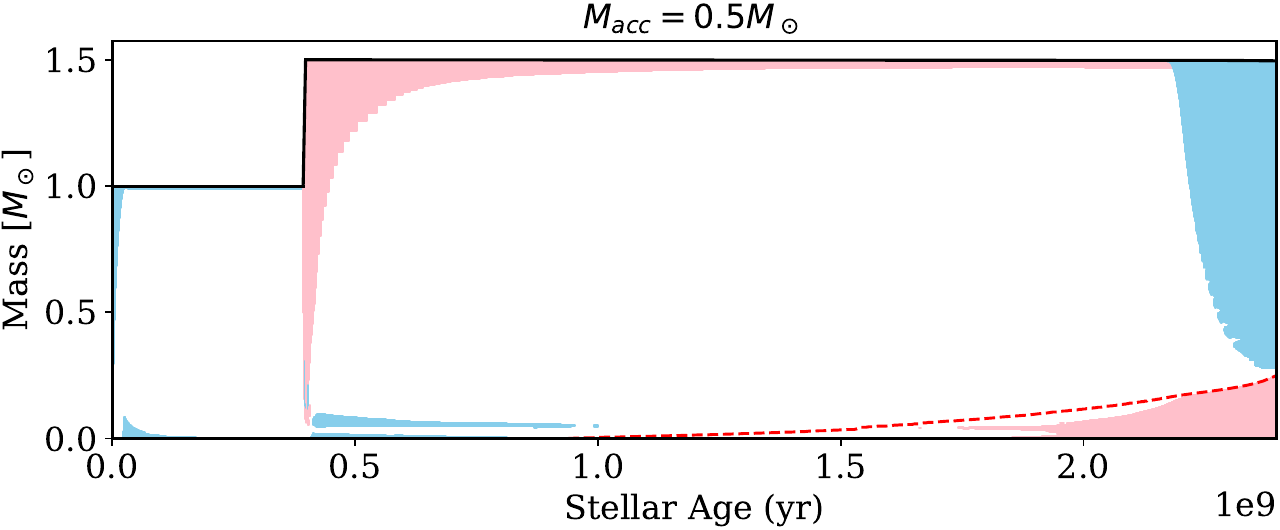}
        \caption{}
        \label{Kipp0.5}
    \end{subfigure}\\

    \caption{Kippenhahn diagram for a Ba star with final mass of $1.5~\rm{M}_\odot$, Z=0.007 model accreting (\textit{panel a}) $0.05~\rm{M}_\odot$, (\textit{panel b}) $0.3~\rm{M}_\odot$, and (\textit{panel c}) $0.5~\rm{M}_\odot$ from a $3~\rm{M}_\odot$, Z=0.007 AGB star. The regions of TH mixing and convection are shown in pink and blue, respectively. The mass of the He-core is shown with the dotted red line. The depth of the TH mixing region increases with an increase in the accreted mass.}
    \label{Kipp_acc}
\end{figure}

Figure~\ref{Kipp_acc} shows the Kippenhahn diagram for a Ba star of final mass $1.5~\rm{M}_\odot$, accreting $0.05~\rm{M}_\odot$, $0.3~\rm{M}_\odot$, and $0.5~\rm{M}_\odot$ from a $3~\rm{M}_\odot$, Z=0.007 AGB star. The TH mixing region that develops post-accretion is shown in pink, and convection is shown in blue. The depth of the TH mixing region progressively increases from $0.85~\rm{M}_\odot$ when the star accretes $0.05~\rm{M}_\odot$ and approaches the stellar core when a substantial mass of $0.5~\rm{M}_\odot$ is accreted.

\end{appendix}

\end{document}